\documentstyle[iopfts]{ioplppt}
\eqnobysec
\begin{document}
\jl{19}
\title{Excitation theory for space-dispersive active media
           waveguides}[Excitation theory for SDAM waveguides]

\author{Anatoly A. Barybin}
\address{Electronics Department, Electrotechnical University,
                               St. Petersburg, 197376, Russia}
\begin{abstract}

A unified electrodynamic approach to the guided-wave excitation theory
is generalized to the waveguiding structures containing a hypothetical
space-dispersive medium with drifting charge carriers possessing
simultaneously elastic, piezoelectric and magnetic properties. Substantial
features of our electrodynamic approach are: (i) the allowance for medium
losses and (ii) the separation of potential fields peculiar to the slow
quasi-static waves which propagate in such active media independently of
the fast electromagnetic waves of curl nature. It is shown that the
orthogonal complementary fields appearing inside the external source
region are just associated with a contribution of the potential fields
inherent in exciting sources. Taking account of medium losses converts
the usual orthogonality relation into a novel form called the
quasi-orthogonality relation. Development of the mode quasi-orthogonality
relation and the equations of mode excitation is based on the generalized
reciprocity relation (the extended Lorentz lemma) specially proved for
this purpose with allowing for specific properties of the space-dispersive
active media and separating the potential fields. The excitation equations
turn out to be the same in form whatever waveguide filling, including both
the time-dispersive (bianisotropic) and space-dispersive media. Specific
properties of such media are reflected in a particular form of the
normalizing coefficients for waveguide eigenmodes. It is found that the
separation of potential fields reveals the fine structure of interaction
between the exciting sources and mode eigenfields: in addition to the
exciting currents (bulk and surface) interacting with the curl fields, the
exciting charges (bulk and surface) and the double charge (surface dipole)
layers appear to interact with the quasi-static potentials and the
displacement currents, respectively.

\end{abstract}
\maketitle

\section{Introduction}

Modern progress of material science and technology opens new potential
possibilities in synthesizing complex and composite media with unique
electromagnetic properties at microwaves and in optics. This requires
revising some problems of guided-wave electrodynamics, in particular,
the theory of waveguide excitation by external sources. The guided-wave
excitation theory for passive media with isotropic, anisotropic and
bi-anisotropic properties was developed in [1].
Electrodynamic processes in such passive media are fully described
by Maxwell's equations and local constitutive relations, the most
general form of which is inherent in bianisotropic media (BAM) with
frequency-dependent constitutive parameters. Although magneto-electric
phenomena in such media, by their microscopic nature, are brought about
by non-locality of short-range polarization responses on electromagnetic
actions~[2\,--\,4], their macroscopic manifestations are actually similar
to those for real time-dispersive media. Indeed, the short-range character
of these phenomena enables one to use the plane wave representation with
wavenumber $k=\omega/c$ so that in the first-order approximation all the
constitutive tensor parameters of such a medium become solely local and
frequency-dependent (see Ref.\ [2]).

Unlike passive media, a true active medium requires, in addition to
Maxwell's equations, an appropriate equation of motion for
its electrodynamic description. Among {\it active media\/} we restrict
our consideration to three kinds:\\
(i) piezo-dielectrics with elastic properties providing the
technological basis for acoustic-wave electronics~[5\,--\,7],\\
(ii) dielectrics with ferrimagnetic properties (magnetized ferrites)
forming the technological basis for spin-wave electronics~[8\,--\,10],\\
(iii) nondegenerate plasmas with drifting charge carriers (in
particular, semiconductors with negative differential mobility of hot
electrons) constituting the technological basis for plasma-wave
electronics [11\,--\,14].

For generality, we shall investigate a hypothetical space-dispersive
medium possessing simultaneously elastic, piezoelectric, ferrimagnetic
and nondegenerate plasma properties in order to provide specific relations
for any kind of the complex composite medium as a special case of the
general situation developed below.

Space-dispersive properties of such a waveguiding medium are related to
non-local effects caused by specific interactions between adjacent
particles of the active media, such as elastic interactions in
piezo-dielectrics, exchange interactions in ferrites and carrier
diffusion effects in nondegenerate plasmas. Neglecting the non-local
effects enables the equations of medium motion to be converted into the
constitutive relations with frequency-dependent parameters.
In other words, such a medium possesses properties of the usual
{\it time-dispersive active medium\/} (TDAM)
and in this case its electrodynamic properties fully conform to the
ordinary anisotropic media examined in~[1]. Our subsequent
analysis will be based on results obtained there, among them a novel
notion of the mode quasi-orthogonality for lossy waveguides (see
section 3 of~[1]).

The objective in writing the paper is to generalize the guided-wave
excitation theory developed previously for the waveguiding structures
with time-dispersive bianisotropic media~[1] to waveguides filled with
a generalized (hypothetical) space-dispersive medium which contains all
the piezoelectric, ferrimagnetic and plasma phenomena. Section~2 includes
input information about the modal field expansions with separating
potential fields to give a physical insight into the nature of the
orthogonal complementary fields obtained previously~[1]. Since the
electrodynamic description of space-dispersive phenomena requires, in
addition to Maxwell's equation, special equations of medium motion,
section~3 is devoted to consideration of the appropriate equations for
three kinds of SDAM. Section~4 is fundamental and begins with an examination
of the generalized reciprocity relation specially derived in the Appendix.
This relation serves as a basis for obtaining the mode quasi-orthogonality
relation and the equations of mode excitation taking into account specific
contributions from the generalized hypothetical medium. Section~5 contains
the general discussion of physical features in the mathematical
description concerning the excitation of lossless and lossy systems valid
for both BAM (see [1]) and SDAM waveguides.
Mathematical notation here is the same as applied in~[1],
in particular, $A$ stands for scalars, $\bi A$ for vectors,
$\bar{\!\bi A}$ for dyadics and $\bar{\bar{\!\bi A}}$ for tensors
of rank more than two.

\section{Modal field expansions with separating potential fields}
\label{sec:1}

The fundamental result of the previous examination~[1], which was
obtained for passive media (including BAM as the most general case),
is the incompleteness of eigenmode basis for any waveguiding structure
inside the excitation source region.
This manifests itself in the fact that the modal field expansions
${\bi E}_a=\sum_k A_k{\bi E}_k$ and ${\bi H}_a=\sum_k A_k{\bi H}_k$ must
be supplemented with the orthogonal complementary fields ${\bi E}_b$ and
${\bi H}_b$ which are related to the longitudinal components of the
external currents ${\bi J}_{\rm b}^{\rm e}$ and
${\bi J}_{\rm b}^{\rm m}$,\, i.e.\ the complete electromagnetic fields
inside the source region are represented in the following form
(cf.\ equations~(4.8) and (4.9) in~[1])
\begin{equation}  \fl
{\bi E}({\bi r}_{\rm t},z) = {\bi E}_a({\bi r}_{\rm t},z) +
{\bi E}_b({\bi r}_{\rm t},z) =
\sum_k A_k(z)\,{\bi E}_k({\bi r}_{\rm t},z) +
{\bi E}_b({\bi r}_{\rm t},z)
\label{eq:1.1}
\end{equation}
\begin{equation}   \fl
{\bi H}({\bi r}_{\rm t},z)\!= {\bi H}_a({\bi r}_{\rm t},z) +
{\bi H}_b({\bi r}_{\rm t},z) =
\sum_k A_k(z)\,{\bi H}_k({\bi r}_{\rm t},z) +
{\bi H}_b({\bi r}_{\rm t},z)
\label{eq:1.2}
\end{equation}
where an unknown longitudinal dependence of the excitation amplitudes
$A_k(z)$ is due to exciting external sources as well as the
complementary fields ${\bi E}_b$ and ${\bi H}_b$.

In~[1] the complementary fields received a mathematical
substantiation as the orthogonal complement to Hilbert space spanned
by the eigenfield basis $\{{\bi E}_k,\,{\bi H}_k\}$, but their physical
nature is still not properly understood. The present examination of
space-dispersive media enables us to furnish an explanation of these
fields as a part of the potential fields generated by external sources.

The basic electrodynamic property of active media is associated with their
ability to support propagating the special kind of slow waves (whose
propagation velocity is much less than velocity of light characteristic
of a medium under consideration) such as the surface acoustic waves (SAW)
in elastic piezo-dielectrics~[5\,--\,7], the magnetostatic spin waves (MSW)
in magnetized ferrites~[8\,--\,10] and the space-charge waves (SCW) in
semiconductors with negative differential mobility of electrons
[12\,--\,16]. These slow waves, being of multimodal character for
composite (multilayered) structures, constitute the so-called quasi-static
part of the total electromagnetic spectrum, whose principal feature is
related to the predominance of a relevant potential field (electric for
SAW and SCW or magnetic for MSW) over its curl counterpart.

Representation of the total fields $({\bi E},\,{\bi H})$ as a sum of their
curl $({\bi E}_{\rm c},\,{\bi H}_{\rm c})$  and potential
$({\bi E}_{\rm p}\!=\!-\bnabla\varphi,\;{\bi H}_{\rm p}\!=
\!-\bnabla\psi)$ parts is realizable on the basis of
Helmholtz's decomposition theorem~[17]. In this theorem, the fields of
every $k$th eigenmode can be represented as
\begin{equation}      \fl
{\bi E}_k =\,{\bi E}_{{\rm c}k} + {\bi E}_{{\rm p}k} =\,
{\bi E}_{{\rm c}k} -\bnabla\varphi_k              \qquad
{\bi H}_k =\,{\bi H}_{{\rm c}k} + {\bi H}_{{\rm p}k} =\,
{\bi H}_{{\rm c}k} -\bnabla\psi_k
\label{eq:1.3}
\end{equation}
where $\bnabla\bdot{\bi E}_{{\rm c}k}= 0$ \,and\,
$\bnabla\bdot{\bi H}_{{\rm c}k}= 0$.

Therefore, instead of one set of the total eigenfields $\{{\bi E}_k,\,
{\bi H}_k\}$ forming the basis of Hilbert space, we now have
two sets involving the curl eigenfields $\{{\bi E}_{{\rm c}k},\,
{\bi H}_{{\rm c}k}\}$
and the quasi-static eigenpotentials $\{\varphi_k,\,\psi_k\}$. This
extends dimensionality of Hilbert space and allows us to
anticipate the possibility of expanding the complementary fields
${\bi E}_b$ and ${\bi H}_b$ in terms of the scalar potential basis
$\{\varphi_k,\,\psi_k\}$.

Let us apply the vector curl-field basis $\{{\bi E}_{{\rm c}k},\,
{\bi H}_{{\rm c}k}\}$ to expand the desired curl fields
\begin{equation}     \fl
{\bi E}_{\rm c}({\bi r}_{\rm t},z) =
\sum_k A_k(z)\,{\bi E}_{{\rm c}k}({\bi r}_{\rm t},z)      \qquad
{\bi H}_{\rm c}({\bi r}_{\rm t},z) =
\sum_k A_k(z)\,{\bi H}_{{\rm c}k}({\bi r}_{\rm t},z)
\label{eq:1.4}
\end{equation}
and the scalar potential basis $\{\varphi_k,\,\psi_k\}$ to expand the
desired quasi-static potentials
\begin{equation}     \fl
\varphi({\bi r}_{\rm t},z)= \sum_k A_k(z)\,\varphi_k({\bi r}_{\rm t},z)
\qquad
\psi({\bi r}_{\rm t},z)= \sum_k A_k(z)\,\psi_k({\bi r}_{\rm t},z)\,.
\label{eq:1.5}
\end{equation}

In this case the complete fields inside the source region can be written,
allowing for relations~(\ref{eq:1.3}), in the following form
\begin{eqnarray}
{\bi E}\,=\,{\bi E}_{\rm c}\,- \bnabla\varphi &=
\sum_k A_k\,({\bi E}_{{\rm c}k} -\!
\bnabla\varphi_k)\,-\,
{\bi z}_0 \sum_k {dA_k\over dz}\,\varphi_k   \nonumber\\
&= \sum_k A_k\,{\bi E}_k \,-\,
{\bi z}_0 \sum_k {dA_k\over dz}\,\varphi_k
\label{eq:1.6}
\end{eqnarray}
\begin{eqnarray}
{\bi H} =\,{\bi H}_{\rm c}\,- \bnabla\psi &=
\sum_k A_k\,({\bi H}_{{\rm c}k} -\!
\bnabla\psi_k)\,-\,
{\bi z}_0 \sum_k {dA_k\over dz}\,\psi_k      \nonumber\\
&= \sum_k A_k\,{\bi H}_k \,-\,
{\bi z}_0 \sum_k {dA_k\over dz}\,\psi_k .
\label{eq:1.7}
\end{eqnarray}
From comparison of equations~(\ref{eq:1.6}) and (\ref{eq:1.7}) with
(\ref{eq:1.1}) and (\ref{eq:1.2}) it follows that
\begin{equation}
{\bi E}_b = -\,{\bi z}_0 \sum_k {dA_k\over dz}\,\varphi_k
\qquad
{\bi H}_b = -\,{\bi z}_0 \sum_k {dA_k\over dz}\,\psi_k .
\label{eq:1.8}
\end{equation}

Formulae~(\ref{eq:1.8}) give the expected expansions
of complementary fields in terms of the quasi-static potentials of
eigenmodes in a specific form involving the derivatives $dA_k/dz$
in place of the amplitudes $A_k$ as expansion coefficients, which vanish
outside the source region where $A_k(z)$ = const. As evidently follows
from (\ref{eq:1.6}) and (\ref{eq:1.7}), the complementary fields
(\ref{eq:1.8}) are in fact a part of the total potential
fields associated with external sources, which was formerly unexpandable,
whereas the other part is included in the modal expansions.

All the above results lead us to the important conclusion that the
complemented Hilbert space spanned by two sets of base functions,
consisting of the curl eigenfields $\{{\bi E}_{{\rm c}k},\,
{\bi H}_{{\rm c}k}\}$ and the quasi-static eigenpotentials
$\{\varphi_k,\, \psi_k\}$, is closed
with respect to any function corresponding to arbitrary external
sources because the desired representations for the curl
fields~(\ref{eq:1.4}) and for the quasi-static potentials~(\ref{eq:1.5})
do not contain any orthogonal complements. So, if we entirely exclude the
potential fields ${\bi E}_{\rm p}$ and ${\bi H}_{\rm p}$ from our analysis,
by using instead their scalar potentials $\varphi$ and $\psi$,
together with the curl fields ${\bi E}_{\rm c}$ and ${\bi H}_{\rm c}$,
\,the appropriate sets of the mode quantities
$\{\varphi_k,\,\psi_k\}$ and $\{{\bi E}_{{\rm c}k},\,
{\bi H}_{{\rm c}k}\}$ will constitute a complete basis that
produces the modal expansions (\ref{eq:1.4}) and (\ref{eq:1.5}) with no
orthogonal complements. In consequence, the latter fact provides the
disappearance of the effective surface currents
${\bi J}_{\rm s,ef}^{\rm e}$ and ${\bi J}_{\rm s,ef}^{\rm m}$ created
by the complementary fields and given by formula~(4.32) in~[1].

In the subsequent examination, we shall consider that the modal expansions
(\ref{eq:1.4}) and (\ref{eq:1.5}) involve the transverse distributions of
physical quantities for all the modes of SDAM waveguide, which are taken
to be known from a preliminary solution to the appropriate source-free
boundary-value problem. Hence, the basic task is to find the mode
excitation amplitudes $A_k(z)$ inside the external source region.

\section{Constitutive relations and equations of motion for SDAM}
\label{sec:2}

As was mentioned above, we consider the generalized (hypothetical) medium
possessing simultaneously piezoelectrically-elastic, ferrimagnetic and
plasma properties and restrict our consideration to the macroscopic model
of SDAM, i.e.\ the medium under examination is regarded as a continuum
characterized by pertinent phenomenological parameters. Such is the case
in the long-wavelength approximation, when an excitation wavelength is much
greater than typical medium dimensions such as interatomic distances for
solids or Debye length for plasmas. For hot electrons in non-degenerate
semiconductor plasmas, this holds true for the hydrodynamic,
quasi-hydrodynamic and local-field approximations [14].

\subsection{Piezoelectrically-elastic properties of a medium}
\label{sec:2A}

The stressed state of an elastic medium is specified by two second-rank
tensors: the stress tensor $\bar{\!\bi T}$ and the strain tensor
$\bar{\!\bi S}$. The components $S_{ij}$ are related to components $u_i$
of the vector ${\bi u}$ of medium particle displacement by the known
relation~[5\,--\,7]
\begin{equation}
S_{ij} = {1\over2}\, \biggl( {\partial u_i\over\partial r_j} +
{\partial u_j\over\partial r_i} \biggr) \,.
\label{eq:2.8}
\end{equation}

The components $T_{ij}$ of stress tensor enter into the dynamic equation
for the elastic medium written in the ordinary form of
Newton's equation~[5\,--\,7]
\begin{equation}
\rho_{\rm m}{\d U_i\over\d t}\,=\,{\partial T_{ij}\over\partial r_j}
\qquad \mbox{or} \qquad
\rho_{\rm m}{\d{\bi U}\over\d t}\,=\,
\bnabla\bdot\bar{\bi T}
\label{eq:2.9}
\end{equation}
where $\rho_{\rm m}$ is the mass density and ${\bi U} =
{\partial{\bi u}/\partial t}\equiv\dot{\bi u}$\, is the medium particle
velocity. The left part of equation~(\ref{eq:2.9}) involves the total
derivative with respect to time \,$\d{\bi U}/\d t =
{\partial{\bi U}/\partial t}+{\bi U}\bdot\bnabla{\bi U}$
and the right part expresses the dynamic force ${\bi F}=\!
\bnabla\bdot\bar{\!\bi T}$ exerted on a unit mass
element (still without allowing for dissipative effects).

If the elastic dielectric medium possesses piezoelectricity,
both the strain $\bar{\!\bi S}$ and the electric field ${\bi E}$ evoke
an appearance of the electric polarization ${\bi P}$ and the elastic
stress $\bar{\!\bi T}$. In this case the constitutive relations can be
written in one of the conventional forms (ignoring magnetostrictive
effects)~[5\,--\,7]
\begin{eqnarray}
P_k =
e_{kij}\,S_{ij} + \epsilon_0\,\chi_{ik}^{\,\scriptscriptstyle S} E_i
\qquad &\mbox{or}& \qquad
{\bi P} =\,
\bar{\bar{\!\bi e}}:\bar{\bi S}\,+ \epsilon_0\,\,
\bar{\!\bchi}^{\,\scriptscriptstyle S}\!
\bdot{\bi E}
\label{eq:2.10}\\
T_{ij}\,=
c_{ijkl}^{\,\scriptscriptstyle E}\,S_{kl} - e_{kij} E_k
\qquad &\mbox{or}& \qquad
\bar{\!\bi T} =\;
\bar{\bar{\!\bi c}}^{\,\scriptscriptstyle E}\!:
\bar{\bi S}\,-\, \bar{\bar{\!\bi e}}\bdot{\bi E} .
\label{eq:2.11}
\end{eqnarray}
In view of the relation ${\bi D}= \epsilon_0{\bi E}+ {\bi P}$\,
(\ref{eq:2.10}) can be rewritten as
\begin{equation}
D_k =
e_{kij}\,S_{ij}\,+\,\epsilon_{ik}^{\,\scriptscriptstyle S}\,E_i
\qquad \mbox{or} \qquad
{\bi D} =\,
\bar{\bar{\!\bi e}}:\bar{\!\bi S}\,+\;
\bar{\!\bepsilon}^{\,\scriptscriptstyle S}\!
\bdot{\bi E}\,.
\label{eq:2.12}
\end{equation}

In (\ref{eq:2.10}) -- (\ref{eq:2.12}), the quantities
\,$\bar{\!\bchi}^{\,\scriptscriptstyle S}$ and\,
$\bar{\!\bepsilon}^{\,\scriptscriptstyle S}\!= \epsilon_0
\,(\bar{\bi I}+\,\bar{\!\bchi}^{\,\scriptscriptstyle S})$
are the second-rank susceptibility and permittivity tensors,\,
$\bar{\bar{\!\bi e}}$ is the third-rank piezoelectric stress
tensor and\, $\bar{\bar{\!\bi c}}^{\,\scriptscriptstyle E}$
is the fourth-rank elastic stiffness tensor (for brevity sake, the
superscripts $S$ and $E$ will be dropped below).

Acoustic-wave propagation losses in solids are caused by two dissipative
effects which can be introduced into the dynamic equation~(\ref{eq:2.9})
phenomenologically by means of the following quantities:

(i) the {\it internal friction\/} stress \,$\bar{\!\bi T}^{\rm fr}$
associated with the existence of viscous properties of an elastic medium,
which should be added to the stress tensor $\bar{\!\bi T}$ (to yield
the total stress tensor $\bar{\bi T}^{\scriptscriptstyle\Sigma}=
\bar{\bi T}+ \bar{\bi T}^{\rm fr}$) in the form analogous to that used
for an isotropic medium~[7]
\begin{equation}
T_{ij}^{\rm fr}\!=\,\eta_{ijkl}\,{\partial S_{kl}\over\partial t}
\qquad \mbox{or} \qquad
\bar{\!\bi T}^{\rm fr}\!=\,\;\bar{\bar{\!\bfeta}}:
\dot{\bar{\!\bi S}}
\label{eq:2.13}
\end{equation}
where \,$\bar{\bar{\!\bfeta}}$ is the viscosity
tensor considered as phenomenologically given;

(ii) the {\it dynamic friction\/} force ${\bi F}^{\rm fr}$ exerted by
imperfections of a crystal lattice on the motion of acoustic phonons,
which should be added to the total dynamic force
${\bi F}^{\scriptscriptstyle\Sigma}=
\bnabla\!\bdot\bar{\bi T}^{\scriptscriptstyle\Sigma}$
in the form of a relaxation term
\begin{equation}
F_i^{\rm fr} = -\,\tau_{ij}^{-1} \rho_{\rm m} U_i
\qquad \mbox{or} \qquad
{\bi F}^{\rm fr} =
-\;\bar{\!\btau}^{-1}\!\bdot\rho_{\rm m}{\bi U}
\label{eq:2.14}
\end{equation}
where \,$\bar{\!\btau}^{-1}$ is the inverse relaxation
time tensor regarged as phenomenologically given.

Allowing for relations (\ref{eq:2.13}) and (\ref{eq:2.14}),
equations~(\ref{eq:2.8}) and (\ref{eq:2.9}) are rewritten
in the following form
\begin{equation}
{\partial S_{ij}\over\partial t} =\,
{1\over2}\, \biggl( {\partial U_i\over\partial r_j} +
{\partial U_j\over\partial r_i} \biggr)
\label{eq:2.15}
\end{equation}
\begin{equation}
\rho_{\rm m} {\d{\bi U}\over\d t}\,=\,
\bnabla\bdot\bar{\bi T}^{\scriptscriptstyle\Sigma}\,+\,{\bi F}^{\rm fr}\,.
\label{eq:2.16}
\end{equation}

In the case of pure harmonic processes (with time dependence in the
form of $\exp(\i\omega t)$), the above equations can be linearized
in small-signal quantities (marked by subscript~1 unlike their static
values marked by subscript~0) so that \,$\rho_{\rm m}=\rho_{{\rm m}0}+
\rho_{{\rm m}1},\, {\bi U}={\bi U}_1$, etc\, with $|\rho_{{\rm m}1}|\ll
|\rho_{{\rm m}0}|$.

\subsection{Ferrimagnetic properties of a medium}
\label{sec:2B}

Macroscopic dynamics of a ferrimagnetic medium uniformly magnetized by an
external static field ${\bi H}_0^{\rm e}$ to the saturation magnetization
${\bi M}_0$ is described by the equation of motion written for the total
magnetization vector $\bi M$ in the following form~[8\,--\,10]
\begin{equation}
{\partial{\bi M}\over\partial t}= -\,\gamma\mu_0\,
({\bi M}\times{\bi H}_{\rm eff})\,+\,{\bi R}
\label{eq:2.21}
\end{equation}
where $\gamma=|e|/m_0$ is the gyromagnetic ratio  for magnetism of
spin nature.

The effective magnetic field \,${\bi H}_{\rm eff}$\, takes into
consideration all the torque-producing contributions caused, in addition
to ${\bi H}_0^{\rm e}$, \,by:\, (a) the Maxwellian field ${\bi H}$
(satisfying Maxwell's equations),\, (b) the crystal anisotropy field
${\bi H}_{\rm c}=-\;\bar{\bcal N}_{\rm c}\bdot\,{\bi M}$ (due to
magnetocrystalline anisotropy of a ferrite material),\,
(c) the demagnetizing field ${\bi H}_{\rm d} =
-\;\bar{\bcal N}_{\rm d}\bdot\,{\bi M}$ (due to shape
anisotropy of a ferrite sample),\, (d) the exchange field
${\bi H}_{\rm ex}= \lambda_{\rm ex}\bnabla^2{\bi M}$
(due to nonuniform exchange interaction of precessing spins),
namely~[8\,--\,10]
\begin{equation}
{\bi H}_{\rm eff}\,=\,
{\bi H}_0^{\rm e}\,+\,{\bi H}\,-\;
\bar{\bcal N}\bdot{\bi M}\,+\,
\lambda_{\rm ex}\bnabla^2{\bi M} \,.
\label{eq:2.22}
\end{equation}
Here the net anisotropy tensor \,$\bar{\bcal N}\!=\!
\bar{\bcal N}_{\rm c}+ \bar{\bcal N}_{\rm d}$
allowing for both the magnetocrystalline anysotropy of a medium and
the demagnetization anisotropy of a ferrite sample is assumed to be
known, as well as the exchange constant $\lambda_{\rm ex}$.

The relaxation term ${\bi R}$ taking account of magnetic losses in
ferrites is written in different forms, among them more
convenient for us is the Gilbert form~[8\,--\,10]
\begin{equation}
{\bi R}= \alpha\,\biggl(
{{\bi M}\over M_0}\times{\partial{\bi M}\over\partial t} \biggr)
\label{eq:2.23}
\end{equation}
with the damping parameter $\alpha$ considered as phenomenologically
given or found from the resonance line half-width measurements as
$\alpha=\Delta H/H_0$~[10].

The Maxwellian field ${\bi H}$, by its sense, is always a signal quantity,
i.e.\ ${\bi H}\equiv{\bi H}_1$, unlike the total magnetization $\bi M$
which is represented by separating a small-signal magnetization
${\bi M}_1$ in the form ${\bi M}({\bi r},t)={\bi M}_0+{\bi M}_1({\bi r},t)$
where $|{\bi M}_1|<<|{\bi M}_0|$. Then the effective magnetic field
(\ref{eq:2.22}) takes the following form
\begin{equation}
{\bi H}_{\rm eff}\,=\,{\bi H}_0\,+\,
{\bi H}_1\,-\,\bar{\bcal N}\bdot{\bi M}_1\,+\,
\lambda_{\rm ex}\bnabla^2{\bi M}_1
\label{eq:2.33}
\end{equation}
where ${\bi H}_0= {\bi H}_0^{\rm e} -\,\bar{\bcal N}
\bdot{\bi M}_0$ is the static field inside a ferrite sample.

\subsection{Drifting charge carriers in a medium}
\label{sec:2C}

For the hydrodynamic description of nondegenerate plasmas with drifting
streams of mobile charge carriers, it is more suitable and even necessary
to apply, instead of the widely-used Eulerian description, a less known
polarization description [13,\,14,\,18,\,19].

As known~[11\,--\,14], in the hydrodynamic model of non-degenerate
plasmas the drifting charge carriers (say, electrons) are
represented as a charged fluid flow characterized by such macroscopic
quantities as the mean electron density $n$ (or the charge density
$\rho=en$), the mean electron velocity ${\bi v}$ (or the current
density ${\bi J}=\rho{\bi v}$) and the electron temperature $T$
(or the electron pressure $p=nk_BT$). Microscopic processes of scattering
and thermal chaotic motion (or diffusion) of carriers are described in
this model by such phenomenological parameters as the momentum relaxation
time (or the mean time of free path) $\tau$ and the thermal velocity
$v_{\scriptscriptstyle T}=(k_BT/m)^{1/2}$ (or the difffuson constant
${\cal D}= v_{\scriptscriptstyle T}^2\tau$).
When intercarrier (electron-electron) collisions are rather frequent,
there is local thermal equilibrium inside the carrier ensemble
with the electron temperature $T$ exceeding a lattice temperature
$T_0$ for high electric fields.

In approximation of the local thermal equilibrium, the hydrodynamic
force equation has the following form~[11\,--\,14]
\begin{equation}
{\partial{\bi v}\over\partial t}\,+\,
({\bi v}\bdot\bnabla){\bi v}\,=\,
{e\over m}\,({\bi E}\,+\,{\bi v}\times{\bi B})\,-\,
{\bnabla(nk_BT)\over mn}\,-\,{{\bi v}\over\tau}
\label{eq:2.35}
\end{equation}
where $m$ is the effective electronic mass different from the mass
$m_0$ of a free electron.

Equation~(\ref{eq:2.35}) holds true for the case of uniform stationary
heating of electrons when the electron viscosity, heat flow and thermal
perturbations in an electron ensemble are negligibly small, so that
there is no contribution from the so-called thermoforce and
$\bnabla p =mv_{\scriptscriptstyle T}^2\bnabla n$~[14]. Such a
situation takes place when $\tau_{\rm e}<<\tau_{\scriptscriptstyle M}$,\,
where $\tau_{\rm e}$ is the energy relaxation time determining the rate
of electron temperature perturbations and $\tau_{\scriptscriptstyle M}=
\epsilon/\sigma=\epsilon m/e^2n\tau$ is the Maxwellian relaxation time
determining the time scale of ac changes in the electric field and charge
distribution. This condition means that the temperature keeps pace with
signal perturbations in the electric field providing a local relationship
between $T$ and $E$~[14]. The latter allows the momentum relaxation time
$\tau$ to be considered as a function of the electric field magnitude $E$,
which is given phenomenologically or found from measuring the field
dependence of mobility $\mu(E)=(e/m)\tau(E)$ and diffusion constant
${\cal D}(E)=v_{\scriptscriptstyle T}^2\,\tau(E)$.

Strictly speaking, for plasmas placed in a magnetic field ${\bi B}$
the electron heating is produced by an electric field known as the
effective heating field~[14]
\[
E_h =\,\sqrt{ {E^2+ ({\bi b}\bdot{\bi E})^2\over 1\,+\,b^2} }
\]
where the quantity ${\bi b}=\mu{\bi B}$ takes into account an influence
of magnetic fields on the heating effect so that the quantities
$\tau,\;\mu$ and $\cal D$\, now depend on $E_h$.

For a small-signal situation when ${\bi E}={\bi E}_0+{\bi E}_1,\;
{\bi B}={\bi B}_0+{\bi B}_1,\; E_h=E_{h0}+E_{h1}$\, and all ac values
(marked by subscript 1) are assumed to be much smaller than their
dc counterparts (marked by subscript 0), we have
\[
\tau(E_h)=\,\tau(E_{h0})\,+\, \biggl(
{d\tau\over dE_h} \biggr)_{E_{h0}} E_{h1}\,\equiv\,\tau_0\,+\,\tau_1
              \\[-0.1cm]
\]
with \,$\tau_0\equiv\tau(E_{h0})$ and
\begin{equation}       \fl
{\tau_1\over\tau_0}= \biggl(
{d\ln\tau\over d\ln E_h} \biggr)_{E_{h0}} {E_{h1}\over E_{h0}}=
(\kappa_0 -1) {E_{h1}\over E_{h0}}=
(\kappa_0 -1) {{\bi F}_0\over E_0}\bdot
{{\bi E}_1 +{\bi v}_0\times{\bi B}_1\over E_0} \,.
\label{eq:2.37}
\end{equation}
Here we have denoted [14]
\begin{equation}
{\bi F}_0 = { (1+b_0^2\,)\, \Bigl[
{\bi E}_0\,+\,({\bi b}_0\bdot{\bi E}_0){\bi b}_0 \Bigr] \over
(1+\kappa_0\,b_0^2\,)\,+\, \Bigl[ (1+b_0^2\,)\,+\,(1-\kappa_0) \Bigr]
({\bi b}_0\bdot{\bi E}_0)^2/E_0^2 }
\label{eq:2.38}
\end{equation}
where ${\bi b}_0\!=\!\mu_{\rm e}{\bi B}_0$ and the {\it anisotropy
coefficient\/} $\kappa_0\!=\!\mu_{\rm d}/\mu_{\rm e}\,(\le 1$,\, with
inequality being true for hot electrons so that $\kappa_0<0$ for negative
differential mobility) is defined as a ratio of the differential
$(\mu_{\rm d})$ and static $(\mu_{\rm e})$ mobilities which are equal
to~[14]
\[
\mu_{\rm d} ={d\,\mu(E)E\over d\,E} \bigg |_{E=E_{h0}}
\;\quad\mbox{and}\qquad
\mu_{\rm e} =\mu(E_{h0})= {e\over m}\,\tau(E_{h0})\equiv
{e\over m}\,\tau_0 \;.
\]

Thus, the force equation (\ref{eq:2.35}) describes dynamics of the
field-charge perturbations in hot electron gases characterized by the
field dependence of momentum relaxation time $\tau(E)$ given
phenomenologically, which is known as the local field approximation~[14].

The polarization $(P)$ description holds an intermediate position
between the well-known Lagrangian $(L)$ and Eulerian $(E)$ descriptions.
To analyze the small-signal processes in drifting carrier streams it is
customary to consider two states of the stream -- unperturbed (without
a signal) and perturbed (with a signal). For the two states in the
$L$-description the motion of a particular group of charges (located
inside a physically infinitesimal volume called the liquid particle) is
described by a time dependence of two radius-vectors ${\bi r}_0(t)$ and
${\bi r}(t)$ appropriate to the unperturbed and perturbed stream. The
fundamental dynamic variable of the polarization description is defined
as a difference in these radius-vectors for two positions of the same
liquid particle caused by signal action~[14,\,18,\,19]:
\begin{equation}
{\bi r}_1({\bi r}_0,t)\,=\,{\bi r}(t)\,-\,{\bi r}_0(t)
\label{eq:2.40}
\end{equation}
which is considered as a function of the unperturbed radius-vector
${\bi r}_0$ and called the {\it electron displacement vector\/}.
Hence, after introducing the displacement vector~(\ref{eq:2.40})
the ``trajectory'' description of electron motion (typical for the
$L$-variables) is replaced by the ``field'' description (typical for the
$E$-variables). Now we deal with a field of the electron displacement
${\bi r}_1({\bi r_0},t)$ which is completely identical to the field of
the lattice particle displacement ${\bi u}({\bi r_0},t)$ applied in
elasticity theory (see section 3.1).

According to the Eulerian and polarization descriptions, the total
instantaneous velocity ${\bi v}({\bi r},t)$\, of a particular group of
charges satisfying the dynamic equation~(\ref{eq:2.35}) is represented
as~[14,\,18,\,19]
\begin{equation}
{\bi v}({\bi r},t)\,=\,{\bi v}_0({\bi r})\,+\,{\bi u}_1({\bi r},t)\,=\,
{\bi v}_0({\bi r}_0)\,+\,{\bi v}_1({\bi r}_0,t)
\label{eq:2.41}
\end{equation}
where the static velocity ${\bi v}_0$ is taken at two positions
(perturbed, $\bi r$, for the $E$-description and unperturbed, ${\bi r}_0$,
for the $P$-description) of the same group of charges. The small-signal
Eulerian $({\bi u}_1)$ and polarization $({\bi v}_1)$ velocities are
defined by relation (\ref{eq:2.41}) at different space points but at
the same point between them there is the following relation:
\,${\bi v}_1= {\bi u}_1+ ({\bi r}_1\bdot\bnabla){\bi v}_0$.

The polarization velocity ${\bi v}_1$ adheres to the equation of
motion in the $P$-variables which is obtained from equation~(\ref{eq:2.35})
in the following form~[14,\,19]
\[         \fl
{\partial{\bi v}_1\over\partial t}\,+\,
({\bi v}_0\bdot\bnabla){\bi v}_1 \,=
\]
\[          \fl
=\,{e\over m} \Bigl(\,
{\bi E}_1\,+\,({\bi r}_1\bdot\bnabla){\bi E}_0\,+\,
{\bi v}_1\times{\bi B}_0\,+\,{\bi v}_0\times{\bi B}_1\,+\,
{\bi v}_0\times
({\bi r}_1\bdot\bnabla){\bi B}_0 \Bigr)       \\[.1cm]
\]
\begin{equation}      \fl
+\;{v_{\scriptscriptstyle T}^2\over\rho_0}\, \Bigl(\,
\rho_0\bnabla
(\bnabla\bdot{\bi r}_1)\,+\,
\bnabla{\bi r}_1\bdot
\bnabla\rho_0 \Bigr) \,-\,
{{\bi v}_1\over\tau_0}\,+\,{{\bi v}_0\over\tau_0}\,{\tau_1 +
({\bi r}_1\bdot\bnabla)\tau_0\over\tau_0}
\label{eq:2.42}
\end{equation}
where the relaxation times $\tau_0$ and $\tau_1$ are given by
equation~(\ref{eq:2.37}). This equation allows for spatial nonuniformity
in dc quantities ${\bi E}_0,\;{\bi B}_0,\;\rho_0$ and $\tau_0$ caused, for
instance, by nonuniform doping of a semiconductor.

Between the polarization variables ${\bi v}_1$ and ${\bi r}_1$ there
is the following relation~[14,\,18,\,19]
\begin{equation}
{\bi v}_1\,=\,{\partial{\bi r}_1\over\partial t}\,+\,
({\bi v}_0\bdot\bnabla){\bi r}_1
\label{eq:2.43}
\end{equation}
which proves to play a role of the continuity equation
\begin{equation}
{\partial\rho_1\over\partial t}\,+\,
\bnabla\bdot{\bi J}_1\,=\,0 \,.
\label{eq:2.44}
\end{equation}

Eulerian densities of charge $\rho_1$ and current ${\bi J}_1=
\rho_1{\bi v}_0+ \rho_0{\bi u}_1$\, are expressed in the $P$-variables
in terms of the {\it electronic polarization vector\/}
${\bi p}_1=\rho_0{\bi r}_1\,$ by the following relations~[14,\,18,\,19]:
\begin{equation}
\rho_1= -\,\bnabla\bdot{\bi p}_1
\label{eq:2.45}
\end{equation}
\begin{equation}
{\bi J}_1= {\partial{\bi p}_1\over\partial t}\,+\,
\bnabla\times({\bi p}_1\times{\bi v}_0) \,.
\label{eq:2.46}
\end{equation}

It is obvious that expressions (\ref{eq:2.45}) and (\ref{eq:2.46})
satisfy the continuity equation~(\ref{eq:2.44}) identically. The charge
and current densities introduced by these formulae are fully the same as
those produced by the motion of an actual dielectric with the polarization
vector ${\bi p}_1$ moving at velocity ${\bi v}_0$~[20]. It is this fact
that has given the name to the polarization description of mobile charges
{\it in vacuo\/} and plasmas.

\subsection{Electromagnetic properties of a medium}
\label{sec:2D}

Electromagnetic fields are described by the usual Maxwell equations
written for small-signal quantities as
\begin{eqnarray}
\bnabla\times{\bi E}_1 =\,
-\,{\partial{\bi B}_1\over\partial t}              \qquad    & &
\bnabla\times{\bi H}_1 =\,
{\partial{\bi D}_1\over\partial t}\,+\,{\bi J}_1
             \nonumber          \\ & \label{eq:2.47} & \\ [-0.2cm]
\bnabla\bdot{\bi D}_1 = \rho_1
                                               \;\;\qquad\qquad & &
\bnabla\bdot{\bi B}_1 =\,0
             \nonumber
\end{eqnarray}
where the densities of charge $\rho_1$ and current ${\bi J}_1$ are
represented in $P$-variables by~(\ref{eq:2.45}) and
(\ref{eq:2.46}). The induction vectors are associated with the
polarization and magnetization vectors by the known relations~[20]
\begin{equation}
{\bi D}_1 =\,\epsilon_0 {\bi E}_1 +\,{\bi P}_1
\qquad\mbox{and}\qquad
{\bi B}_1 =\,\mu_0 ({\bi H}_1 +\,{\bi M}_1) \,.
\label{eq:2.48}
\end{equation}

For the generalized medium under examination, possessing also
piezoelectric and magnetic properties, its polarization ${\bi P}_1$ and
magnetization ${\bi M}_1$ adhere, respectively, to the constitutive
relation (\ref{eq:2.10}) and to the equation of motion (\ref{eq:2.21})
written in a linearized form.

The polarization description of plasmas provides a convenient way to make
an artificial replacement of plasma by an equivalent magneto-dielectric
medium without mobile charges. Indeed, it is known that a real electric
dipole ${\bi p}_1$ moving at velocity ${\bi v}_0$ is perceived, by a fixed
observer, as two immovable dipoles: electric ${\bi p}_1$ and magnetic
${\bi m}_1= {\bi p}_1\times{\bi v}_0$~[20]. Hence, the polarization
description operating with the electronic polarization vector ${\bi p}_1$
allows any charged medium with mobile carriers to be represented as an
equivalent polarized (with ${\bi p}_1$) and magnetized (with ${\bi m}_1$)
medium with no mobile charges. In this case Maxwell's equations
(\ref{eq:2.47}) by inserting expressions~(\ref{eq:2.45}) and
(\ref{eq:2.46}) can be written in the following form
\begin{eqnarray}
\bnabla\times{\bi E}_1 =\,
-\,{\partial{\bi B}_1\over\partial t}             \qquad   & &
\bnabla\times{\bi H}_1^{\rm p} =\,
{\partial{\bi D}_1^{\rm p}\over\partial t}
             \nonumber          \\ & \label{eq:2.49} & \\ [-0.2cm]
\bnabla\bdot{\bi D}_1^{\rm p}=\,0
                                            \;\;\qquad\qquad  & &
\bnabla\bdot{\bi B}_1 =\,0
             \nonumber
\end{eqnarray}
where the fundamental vectors ${\bi E}_1$ and ${\bi B}_1$ remain
unchanged while the vectors ${\bi H}_1$ and ${\bi D}_1$ are replaced
with their equivalent polarization counterparts ${\bi H}_1^{\rm p}$ and
${\bi D}_1^{\rm p}$. The latter are introduced so that
expressions~(\ref{eq:2.48}) would hold their form with the use of these
new vectors, namely
\begin{equation}
{\bi D}_1^{\rm p}=\,
\epsilon_0 {\bi E}_1 +\,{\bi P}_1^{\scriptscriptstyle\Sigma}
\qquad\mbox{and}\qquad
{\bi B}_1=\,\mu_0 ({\bi H}_1^{\rm p}+\,
{\bi M}_1^{\scriptscriptstyle\Sigma})
\label{eq:2.50}
\end{equation}
where the total polarization and magnetization vectors defined as
\begin{equation}
{\bi P}_1^{\scriptscriptstyle\Sigma}=\,
{\bi P}_1 +\,{\bi p}_1
\qquad\mbox{and}\qquad
{\bi M}_1^{\scriptscriptstyle\Sigma}=\,
{\bi M}_1 +\,{\bi m}_1
\label{eq:2.51}
\end{equation}
take into account the contributions from both a crystal lattice
(${\bi P}_1$ and ${\bi M}_1$) and an electron ensemble (\,${\bi p}_1$
and ${\bi m}_1\equiv{\bi p}_1\times{\bi v}_0$). Thus, the new equivalent
field vectors entering into Maxwell's equations~(\ref{eq:2.49}) in the
polarization description are equal to
\begin{equation}
{\bi D}_1^{\rm p}=\,
{\bi D}_1 +\,{\bi p}_1
\qquad\mbox{and}\qquad
{\bi H}_1^{\rm p}=\,
{\bi H}_1 +\,{\bi v}_0\times{\bi p}_1 \,.
\label{eq:2.52}
\end{equation}

Therefore, using Maxwell's equations of the ``dielectric''
form~(\ref{eq:2.49}) enables one to consider all media including plasmas,
as pure dielectric and describe them by applying the equivalent field
vectors~(\ref{eq:2.52}), which, in addition to the lattice polarizations
(electric ${\bi P}_1$ and magnetic ${\bi M}_1$), allow for the electronic
polarization ${\bi p}_1$ due to mobile charge carriers. In this case, the
components of all field vectors, as follows from equations~(\ref{eq:2.49}),
are continuous on the boundaries of drifting carrier streams. Such a feature
makes it more preferable to use the equivalent ``dielectric'' form of
Maxwell's equations (without ${\bi J}_1$ and $\rho_1$) instead of the usual
form (\ref{eq:2.47}) (with ${\bi J}_1$ and $\rho_1$). The point is that the
latter generate the equivalent surface charge \,$\rho_{\rm s}^{\rm eq}\! =
{\bi n}\bdot{\bi p}_1$ and surface current ${\bi J}_{\rm s}^{\rm eq}\! =
({\bi n}\bdot{\bi p}_1){\bi v}_0\,$ on the boundaries of drifting carrier
streams~[14,\,15,\,21],\, which ensures discontinuity in the appropriate
field components.

\section{Orthogonality and quasi-orthogonality of modes and equations
                                 of mode excitation}
\label{sec:3}

\subsection{The generalized reciprocity relation
                        (the extended Lorentz lemma)}
\label{sec:3A}

Up to this point, we have considered the source-free region of a
waveguiding structure with SDAM whose electromagnetic properties are
described by the usual Maxwell equations~(\ref{eq:2.47}) or their
equivalent ``dielectric'' form~(\ref{eq:2.49}). The latter form, which
allows for the electronic polarization of drifting charge carriers,
along with the lattice polarization and magnetization, is more
suitable for subsequent examinations. Inside the region of exciting
bulk sources $({\bi J}_{{\rm b}1}^{\rm e},\;{\bi J}_{{\rm b}1}^{\rm m},\;
\rho_{{\rm b}1}^{\rm e},\; \rho_{{\rm b}1}^{\rm m})$ its curl equations
are written for pure harmonic processes in the form
\begin{equation}
\bnabla\times{\bi E}_1 =
-\,\i\omega{\bi B}_1 -\, {\bi J}_{{\rm b}1}^{\rm m}
\;\;\quad\mbox{and} \;\;\quad
\bnabla\times{\bi H}_1^{\rm p} =\,
\i\omega{\bi D}_1^{\rm p} \,+\; {\bi J}_{{\rm b}1}^{\rm e}
\label{eq:3.1}
\end{equation}
where the electric and magnetic sources obey the continuity equations
\begin{equation}
\i\omega\rho_{{\rm b}1}^{\rm e} \,+\,
\bnabla\bdot{\bi J}_{{\rm b}1}^{\rm e} = \,0
\qquad \mbox{and} \qquad
\i\omega\rho_{{\rm b}1}^{\rm m} \,+\,
\bnabla\bdot{\bi J}_{{\rm b}1}^{\rm m} = \,0 \,.
\label{eq:3.2}
\end{equation}

As follows from~(\ref{eq:2.50}) -- (\ref{eq:2.52}),\, in the equivalent
magneto-dielectric medium without mobile charges, characterized by the total
(lattice and electronic) polarization ${\bi P}_1^{\scriptscriptstyle\Sigma}=
{\bi P}_1+{\bi p}_1$ and magnetization ${\bi M}_1^{\scriptscriptstyle\Sigma}=
{\bi M}_1+{\bi m}_1$, the field-intensity vectors ${\bi E}_1$ and
${\bi H}_1^{\rm p}= {\bi H}_1 +{\bi v}_0\times{\bi p}_1$ produce the
flux-density vectors ${\bi D}_1^{\rm p}$ and ${\bi B}_1$:
\begin{equation}
{\bi D}_1^{\rm p} =\,
\epsilon_0 {\bi E}_1 +\,{\bi P}_1^{\scriptscriptstyle\Sigma}=\,
\epsilon_0\,{\bi E}_1 +\,{\bi P}_1 +\,{\bi p}_1
\label{eq:3.3}
\end{equation}
\begin{equation}
{\bi B}_1 =\,
\mu_0 ({\bi H}_1^{\rm p}+\,{\bi M}_1^{\scriptscriptstyle\Sigma})=\,
\mu_0\,({\bi H}_1 +\,{\bi M}_1)
\label{eq:3.4}
\end{equation}
where the vectors ${\bi P}_1,\; {\bi M}_1$ and ${\bi p}_1$ reflect
physical properties of the medium discussed in section 3.

A basis for deriving the equations of mode excitation is the reciprocity
relation in complex conjugate form (often called the Lorentz lemma)
extended to the generalized medium under consideration with
piezoelectrically-elastic, ferrimagnetic and plasma properties.

To derive the conjugate reciprocity relation it is necessary, in addition
to the given system of equations (marked with subscript~1), to consider
another system (marked with subscript~2 having also small-signal meaning)
whose dynamic equations are all taken in complex conjugate form
(see~(5.25) and (5.26) of~[1]). A conventional procedure applied to these
equations gives relation~(5.27) (see section~5.2.1 in~[1]), which can be
rewritten with the aid of equations~(\ref{eq:3.3})
and (\ref{eq:3.4}) in the following form
\[
\bnabla\bdot \Bigl(
{\bi E}_1\times{\bi H}_2^{{\rm p}*} \,+\,
{\bi E}_2^*\times{\bi H}_1^{\rm p} \Bigr) \,=
-\;\i\omega \Bigl( {\bi P}_1\bdot{\bi E}_2^* -
{\bi P}_2^*\bdot{\bi E}_1 \Bigr)
\]
\[
-\;\i\omega\mu_0 \Bigl( {\bi M}_1\bdot{\bi H}_2^* -
{\bi M}_2^*\bdot{\bi H}_1 \Bigr) \,-\,
\i\omega \Bigl( {\bi p}_1\bdot{\bi E}_2^{\prime*} -\,
{\bi p}_2^*\bdot{\bi E}_1^\prime \Bigr)
\]
\begin{equation}
-\; \Bigl( {\bi J}_{{\rm b}1}^{\rm e}\bdot{\bi E}_2^* \,+\,
{\bi J}_{{\rm b}2}^{{\rm e}*}\bdot{\bi E}_1 \Bigr) \,-\,
\Bigl( {\bi J}_{{\rm b}1}^{\rm m}\bdot{\bi H}_2^{{\rm p}*} \,+\,
{\bi J}_{{\rm b}2}^{{\rm m}*}\bdot{\bi H}_1^{\rm p} \Bigr)
\label{eq:3.5}
\end{equation}
where ${\bi E}_{1,2}^\prime={\bi E}_{1,2}+{\bi v}_0\times{\bi B}_{1,2}$
is the electric field measured relative to an observer moving with
the nonrelativistic velocity ${\bi v}_0$.

The first three terms on the right of equation~(\ref{eq:3.5}) are calculated
in the Appendix by the help of the appropriate equations of medium motion
for elastic piezo-dielectrics, magnetized ferrites and drifting charge
carrier streams (see section 3). Substitution of equations~(\ref{eq:A10}),
(\ref{eq:A14}) and (\ref{eq:A21}) into~(\ref{eq:3.5}) finally
gives the desired reciprocity relation (cf. equation~(5.28) in [1])
\begin{equation}
\bnabla\bdot{\bi S}_{12} +\, q_{12} =\,
r_{12}^{({\rm b})} \,.
\label{eq:3.6}
\end{equation}
Unlike equation~(5.28) of [1], here the total power quantity
\begin{equation}
{\bi S}_{12} =\, {\bi S}_{12}^{\rm EM}\,+\,{\bi S}_{12}^{\rm PM}
\label{eq:3.7}
\end{equation}
in addition to the usual electromagnetic contribution (with the
polarization modification ${\bi H}_{1,2} \to {\bi H}_{1,2}^{\rm p}$
[13,\,21,\,22])
\begin{equation}
{\bi S}_{12}^{\rm EM} =
\Bigl( {\bi E}_1\times{\bi H}_2^{{\rm p}*} \,+\,
{\bi E}_2^*\times{\bi H}_1^{\rm p} \Bigr)
\label{eq:3.8}
\end{equation}
contains also the contribution from non-electromagnetic fields in
polarized media
\begin{eqnarray}
{\bi S}_{12}^{\rm PM} =
&-\,\Bigl( \bar{\bi T}_1^{\scriptscriptstyle\Sigma}\bdot{\bi U}_2^* +
\bar{\bi T}_2^{\Sigma*}\bdot{\bi U}_1 \Bigr)  \nonumber\\
&+\,\Bigl( \bar{\bi V}_1^{\rm m}\bdot{\bi J}_2^{{\rm m}*} +
\bar{\bi V}_2^{{\rm m}*}\bdot{\bi J}_1^{\rm m} \Bigr) \,+\,
\Bigl( \bar{\bi V}_1^{\rm e}\bdot{\bi J}_2^{{\rm e}*} +
\bar{\bi V}_2^{{\rm e}*}\!\bdot{\bi J}_1^{\rm e} \Bigr).
\label{eq:3.9}
\end{eqnarray}
Here we have introduced new quantities:

(i) for ferrimagnetic media -- a vector ${\bi J}_1^{\rm m}$ of the
effective magnetization current density and a tensor
$\bar{\bi V}_1^{\rm m}$ of the effective magnetic (exchange) potential
equal to
\begin{equation}
{\bi J}_1^{\rm m}= \i\omega\mu_0{\bi M}_1   \qquad
\bar{\bi V}_1^{\rm m}= -\,\lambda_{\rm ex}\bnabla{\bi M}_1
\label{eq:2.75}
\end{equation}

(ii) for plasma media -- a vector ${\bi J}_1^{\rm e}$ of the electronic
polarization current density and a tensor $\bar{\bi V}_1^{\rm e}=
\bar{\bi V}_1^{\rm ek}+V_1^{\rm th}\bar{\bi I}$ of the effective
electronic potential involving the electrokinetic potential tensor
$\bar{\bi V}_1^{\rm ek}$ and the thermal (diffusion) potential
$V_1^{\rm th}$  equal to
\begin{equation}
{\bi J}_1^{\rm e}= \i\omega{\bi p}_1   \qquad
\bar{\bi V}_1^{\rm ek}= {m\over e}\,{\bi v}_0{\bi v}_1^{\rm p}  \qquad
V_1^{\rm th}= {m\over e}\,{v_{\scriptscriptstyle T}^2\over\rho_0}\,\rho_1
\label{eq:2.77}
\end{equation}
where ${\bi v}_1^{\rm p}={\bi v}_1 -(e/2m)\,({\bi r}_1\times{\bi B}_0)$
is the resulting small-signal velocity of electrons in the polarization
description allowing for their rotation in the static magnetic field
${\bi B}_0$ with the Larmor angular velocity
$\bomega_L=-\,(e/2m){\bi B}_0$ [21,\,22].

The relaxation processes in the generalized medium are taken into account
by the sum of three contributions:
\[                  \fl
q_{12} =\,
2\,\Bigl( \omega^2\,\bar{\bi S}_2^* \!:\,
\bar{\bar{\!\bfeta}}: \bar{\bi S}_1 +
\rho_{{\rm m}0}\,{\bi U}_2^* \bdot\,
\bar{\!\btau}^{-1}\!\bdot{\bi U}_1 \Bigr) +\,
2\,\nu_{\scriptscriptstyle M}\mu_0 \biggl(
{\omega\over\omega_{\scriptscriptstyle M}} \biggr)^{\!2}
\Bigl( {\bi M_1}\bdot{\bi M}_2^* \Bigr)
\]
\begin{equation}
+\,{1\over\mu_{\rm e}}\, \Bigl(
({\bi v}_1 - \vartheta_1{\bi v}_0)\bdot{\bi J}_2^{{\rm e}*} \,+\,
({\bi v}_2^* - \vartheta_2^*{\bi v}_0)\bdot{\bi J}_1^{\rm e} \Bigr)
\label{eq:3.10}
\end{equation}
where we have introduced (i) for ferrimagnetic media the magnetic
relaxation frequency $\nu_{\scriptscriptstyle M}=
\alpha\,\omega_{\scriptscriptstyle M}= \alpha\gamma\mu_0 M_0$\, and\,
(ii) for plasma media the quantity
\[
\vartheta_1 \equiv\,{\tau_1+ {\bi r}_1\bdot
\bnabla\tau_0\over\tau_0}\,=\,
(\kappa_0 -1)\,{{\bi E}_0\over E_0}\bdot
{{\bi E}_1 +{\bi v}_0\times{\bi B}_1\over E_0}
\]
with the last expression obtained from~(\ref{eq:2.37}) by ignoring
spatial static non-uniformities (so~that $\bnabla\tau_0= 0$)
and assuming that the vector ${\bi B}_0$ is longitudinal (so that
${\bi b}_0\bdot{\bi E}_0= b_0 E_0$\, and \,${\bi F}_0={\bi E}_0$,\, as a
result of~(\ref{eq:2.38}))~[14].

The interaction of the electromagnetic fields with the bulk external
currents is taken into account by the following term in the right-hand
side of (\ref{eq:3.6}):
\begin{equation}
r_{12}^{({\rm b})} = -\,
\Bigl( {\bi J}_{{\rm b}1}^{\rm e}\bdot{\bi E}_2^* \,+\,
{\bi J}_{{\rm b}2}^{{\rm e}*}\bdot{\bi E}_1 \Bigr) \,-\,
\Bigl( {\bi J}_{{\rm b}1}^{\rm m}\bdot{\bi H}_2^{{\rm p}*} \,+\,
{\bi J}_{{\rm b}2}^{{\rm m}*}\bdot{\bi H}_1^{\rm p} \Bigr) \,.
\label{eq:3.11}
\end{equation}

Representation of the electromagnetic field vectors $({\bi E}_{1,2}\,,\,
{\bi H}_{1,2}^{\rm p})$ as the sum of their curl
$({\bi E}_{{\rm c}1,{\rm c}2}\,,\,{\bi H}_{{\rm c}1,{\rm c}2}^{\rm p})$
and potential $(-\bnabla\varphi_{1,2}\,,\,
-\bnabla\psi_{1,2})$ parts, as in~(\ref{eq:1.3}), gives
\[
\bnabla\bdot \Bigl(
{\bi E}_1\times{\bi H}_2^{{\rm p}*} \,+\,
{\bi E}_2^*\times{\bi H}_1^{\rm p} \Bigr) \,=\,
\bnabla\bdot \Bigl[
\Bigl( {\bi E}_{{\rm c}1}\times
{\bi H}_{{\rm c}2}^{{\rm p}*} \,+\,
{\bi E}_{{\rm c}2}^*\times
{\bi H}_{{\rm c}1}^{\rm p} \Bigr)
\]
\[
+\,\Bigl( \varphi_1(\i\omega{\bi D}_2^{\rm p})^* +
\varphi_2^*(\i\omega{\bi D}_1^{\rm p}) \Bigr) \,+\,
\Bigl( \psi_1(\i\omega{\bi B}_2)^* +
\psi_2^*(\i\omega{\bi B}_1) \Bigr) \Bigr]
\]
\[
+\,\Bigl( {\bi J}_{{\rm b}1}^{\rm e}\bdot
\bnabla\varphi_2^* +
{\bi J}_{{\rm b}2}^{{\rm e}*}\bdot
\bnabla\varphi_1 \Bigr) \,+\,
\Bigl( {\bi J}_{{\rm b}1}^{\rm m}\bdot
\bnabla\psi_2^* +
{\bi J}_{{\rm b}2}^{{\rm m}*}\bdot
\bnabla\psi_1 \Bigr)
\]
\begin{equation}
- \Bigl( (\i\omega\rho_{{\rm b}1}^{\rm e})\varphi_2^* +
(\i\omega\rho_{{\rm b}2}^{\rm e})^*\varphi_1 \Bigr) \,-\,
\Bigl( (\i\omega\rho_{{\rm b}1}^{\rm m})\psi_2^* +
(\i\omega\rho_{{\rm b}2}^{\rm m})^*\psi_1 \Bigr)
\label{eq:3.12}
\end{equation}
where we have used the vector identity
$\bnabla\bdot({\bi A}\times\bnabla\phi)=
\bnabla\bdot(\phi\bnabla\times{\bi A})$
and equations~(\ref{eq:3.1}) and (\ref{eq:3.2}).

The use of (\ref{eq:3.12}) in the reciprocity relation
(\ref{eq:3.6}) leaves expressions~(\ref{eq:3.9}) and (\ref{eq:3.10})
for ${\bi S}_{12}^{\rm PM}$ and $q_{12}$ unchanged, but modifies
${\bi S}_{12}^{\rm EM}$ and $r_{12}^{({\rm b})}$, which are now equal to
\[
{\bi S}_{12}^{\rm EM}\,=\,
\Bigl( {\bi E}_{{\rm c}1}\times{\bi H}_{{\rm c}2}^{{\rm p}*} \,+\,
{\bi E}_{{\rm c}2}^*\times{\bi H}_{{\rm c}1}^{\rm p} \Bigr) 
\]
\begin{equation}
+\,\Bigl( \varphi_1(\i\omega{\bi D}_2^{\rm p})^* +
\varphi_2^*(\i\omega{\bi D}_1^{\rm p}) \Bigr) \,+\,
\Bigl( \psi_1(\i\omega{\bi B}_2)^* +
\psi_2^*(\i\omega{\bi B}_1) \Bigr)
\label{eq:3.13}
\end{equation}
\[
r_{12}^{({\rm b})} = -\,\Bigl( {\bi J}_{{\rm b}1}^{\rm e}\bdot
{\bi E}_{{\rm c}2}^* \,+\,
{\bi J}_{{\rm b}2}^{{\rm e}*}\bdot{\bi E}_{{\rm c}1} \Bigr) \,-\,
\Bigl( {\bi J}_{{\rm b}1}^{\rm m}\bdot{\bi H}_{{\rm c}2}^{{\rm p}*} \,+\,
{\bi J}_{{\rm b}2}^{{\rm m}*}\bdot
{\bi H}_{{\rm c}1}^{\rm p} \Bigr)
\]
\begin{equation}
+\,\Bigl( (\i\omega\rho_{{\rm b}1}^{\rm e})\varphi_2^* \,+\,
(\i\omega\rho_{{\rm b}2}^{\rm e})^*\varphi_1 \Bigr) \,+\,
\Bigl( (\i\omega\rho_{{\rm b}1}^{\rm m})\psi_2^* \,+\,
(\i\omega\rho_{{\rm b}2}^{\rm m})^*\psi_1 \Bigr) \,.
\label{eq:3.14}
\end{equation}

Formula~(\ref{eq:3.13}) involves three contributions to the
transferred power from the electromagnetic curl fields,
quasi-electrostatic potential fields and quasi-magnetostatic
potential fields. Formula~(\ref{eq:3.14}) reflects the interactions of
the curl fields with the external bulk currents and the potential fields
with the external bulk charges (surface exciting sources will be
considered later).

The generalized reciprocity relation (or the extended Lorentz lemma) in
the form~(\ref{eq:3.6}) (with equations~(\ref{eq:3.7}) -- (\ref{eq:3.9}),
(\ref{eq:3.10}) and (\ref{eq:3.11}) for the total electromagnetic fields
\,or with equations~(\ref{eq:3.7}), (\ref{eq:3.9}), (\ref{eq:3.10}),
(\ref{eq:3.13}) and (\ref{eq:3.14}) for the fields separated into curl and
potential parts) provides a basis for deriving the mode
quasi-orthogonality relations and the equations of mode excitation in
SDAM waveguides. To this end, the differential form (\ref{eq:3.6}) of the
reciprocity relation should be converted into an integral form.

\subsection{Quasi-orthogonality and orthogonality relations
                                               for SDAM waveguides}
\label{sec:3B}

Let subscripts 1 and 2 in the reciprocity relation~(\ref{eq:3.6})
correspond to two eigenmodes with numbers $k$ and $l$ for a waveguiding
structure with SDAM in the absence of sources, i.e.\
$r_{kl}^{({\rm b})}= 0$. Then integrating equation~(\ref{eq:3.6}) over
the total cross section $S$ of the waveguide and applying the integral
relation~(2.25) in~[1] give
\begin{equation}
{\d P_{kl}(z)\over\d z}\,+\,Q_{kl}(z)\,=\,0
\label{eq:3.15}
\end{equation}
where the complex cross-power (or self-power for $k=l$)
flow transferred jointly by the $k$th and $l$th modes is
\begin{eqnarray}        \fl
P_{kl}(z) =\, P_{kl}^{\rm EM} + P_{kl}^{\rm PM} =\,
{1\over4} \int_S \Bigl( {\bi S}_{kl}^{{\rm EM}*}\!+
{\bi S}_{kl}^{{\rm PM}*} \Bigr) \!\bdot{\bi z}_0\,\d S   \nonumber\\
\fl ={1\over4} \Bigl( N_{kl}^{\rm EM} + N_{kl}^{\rm PM} \Bigr)
A_k^*A_l\,{\rm e}^{-\,(\gamma_k^* \,+\,\gamma_l)z} =
{1\over4}\,N_{kl}\,A_k^*A_l\,{\rm e}^{-\,(\gamma_k^* \,+\,\gamma_l)z}=
{1\over4}\,N_{kl}\,a_k^*a_l
\label{eq:3.16}
\end{eqnarray}
and the complex cross-power (or self-power for $k=l$)
loss dissipated jointly by the $k$th and $l$th modes is
\begin{equation}    \fl
Q_{kl}(z)=\,{1\over4} \int_S q_{kl}^*({\bi r}_{\rm t},z)\,\d S =
{1\over4}\,M_{kl}\,A_k^*A_l\,{\rm e}^{-\,(\gamma_k^* \,+\,\gamma_l)z} =
{1\over4}\,M_{kl}\,a_k^*a_l \,.
\label{eq:3.17}
\end{equation}

The quantities \,${\bi S}_{kl}^{\rm EM},\;{\bi S}_{kl}^{\rm PM}$ and
$q_{kl}$ in these formulas are obtained from~(\ref{eq:3.8}) or
(\ref{eq:3.13}), (\ref{eq:3.9}) and (\ref{eq:3.10}) by replacing
subscripts~1 and 2 with $k$ and $l$, respectively. The quantities $P_{kl}$
and $Q_{kl}$ defined by (\ref{eq:3.16}) and (\ref{eq:3.17}) determine:\\
(i) the total power flow (cf~(2.34) and (2.46) of~[1])
\begin{equation}        \fl
P(z)= \sum_k\sum_l P_{kl}(z) = {1\over4} \sum_k N_k\,|a_k(z)|^2 +
{1\over2}\,{\rm Re} \sum_k\sum_{l>k} N_{kl}\,a_k^*(z)a_l(z)
\label{eq:2.107}
\end{equation}
(ii) the total power loss (cf.\ equations~(2.35) and (2.47) of paper~[1])
\begin{equation}        \fl
Q(z)= \sum_k\sum_l Q_{kl}(z) = {1\over4} \sum_k M_k\,|a_k(z)|^2 +
{1\over2}\,{\rm Re} \sum_k\sum_{l>k} M_{kl}\,a_k^*(z)a_l(z)
\label{eq:2.108}
\end{equation}
where the mode amplitude $a_k$ is related to the excitation amplitude
$A_k$ by formula $a_k= A_k\exp(-\gamma_kz)$ for every $k$th mode
specified by the propagation constant $\gamma_k=\alpha_k+\i\beta_k$ and
the set of cross-section eigenfunctions (marked with hat over them)
$\{\,\hat{\!\bi E}_k({\bi r}_t),\,\hat{\!\bi H}_k({\bi r}_t),\,
{\rm etc}\}$ such that
${\bi E}_k({\bi r}_t,z)=\hat{\!\bi E}_k({\bi r}_t)\exp(-\gamma_kz),\,
{\bi H}_k({\bi r}_t,z)=\hat{\!\bi H}_k({\bi r}_t)\exp(-\gamma_kz)$, etc.

In (\ref{eq:3.16}) -- (\ref{eq:2.108}), following formulae (2.36)
and (2.37) of~[1], we have introduced the normalizing $(N_{kl})$ and
dissipative $(M_{kl})$ coefficients constructed of the cross section
eigenfield vectors. Expression for the normalizing coefficients $(N_{kl}=
N_{kl}^{\rm EM}+ N_{kl}^{\rm PM})$ is given by formula
(\ref{eq:3.51}) or (\ref{eq:3.53}), while the dissipative coefficients
are equal to
\[          \fl
M_{kl} = 2 \int_S \, \biggl[ \Bigl(
\omega^2 (\!\!\hat{\,\;\bar{\!\bi S}_k^*} \!:\,
\bar{\bar{\!\bfeta}}:\hat{\bar{\!\bi S}}_l) +
\rho_{{\rm m}0}\,(\,\hat{\!\bi U}_k^* \!\bdot\,
\bar{\!\btau}^{-1}\!\bdot\hat{\!\bi U}_l) \Bigr) +
\nu_{\scriptscriptstyle M}\mu_0 \biggl(
{\omega\over\omega_{\scriptscriptstyle M}} \biggr)^{\!2}
\Bigl( \hat{\!\bi M}_k^*\bdot\hat{\!\bi M}_l \Bigr)
\]
\begin{equation}
+ \,{1\over2\mu_{\rm e}} \Bigl(
(\hat{\!\bi v}_k^* - \hat{\vartheta}_k^* {\bi v}_0)\bdot
\hat{\!\bi J}_l^{\rm e} +
(\hat{\!\bi v}_l - \hat{\vartheta}_l {\bi v}_0)\bdot
\hat{\!\bi J}_k^{{\rm e}*} \Bigr) \biggr] \d S \,.
\label{eq:2.114}
\end{equation}

Formula~(\ref{eq:3.15}) is in fact the required relation of mode
quasi-orthogonality which reflects an independent transmission and
dissipation of power by any one of mode pairs $(k,\,l)$ in a lossy system
(see section 3.1 of~[1]). The expression of $P_{kl}$ and $Q_{kl}$
in terms of $N_{kl}$ and $M_{kl}$ is written in~(\ref{eq:3.16})
and (\ref{eq:3.17}), owing to formulae (2.41) and (2.42) in~[1].
Substitution of (\ref{eq:3.16}) and (\ref{eq:3.17}) in equation
(\ref{eq:3.15}) yields the desired form of the general
{\it quasi-orthogonality relation\/}
\begin{equation}
(\gamma_k^* + \gamma_l)\,P_{kl} = Q_{kl}   \qquad\mbox{or}\qquad
(\gamma_k^* + \gamma_l)\,N_{kl} = M_{kl}
\label{eq:3.18}
\end{equation}
which turns into the usual orthogonality relation for lossless
waveguiding structures as a special case with $Q_{kl}= M_{kl}= 0$.

Since relations (\ref{eq:3.18}) are completely coincident with the similar
relations (3.6) and (3.7) of~[1], the reasoning given there in
section~3, concerning the quasi-orthogonality of eigenmodes in lossy
waveguides (when $M_{kl}\ne 0$) and the orthogonality of active and
reactive eigenmodes in lossless waveguides (when $M_{kl}= 0$), remains
true for the waveguiding structures with SDAM.

\subsection{Equations of mode excitation for SDAM waveguides}
\label{sec:3C}

To derive the equations of mode excitation, the fields in the reciprocity
relation~(\ref{eq:3.6}) marked by subscript~1 (which will be dropped for
exciting sources) are assumed to be the desired fields excited by the
bulk and surface sources, whereas those marked by subscript~2 are the known
fields of the $k$th mode without sources. Any one of the physical
quantities~$\it\Phi$ (e.g.\ the components of electromagnetic fields,
potentials, polarizations, etc) is written for the $k$th mode~as
\begin{equation}
\it\Phi_k({\bi r}_{\rm t},z) =
\hat{\!\it\Phi}_k({\bi r}_{\rm t})\,{\rm e}^{-\,\gamma_k z}
\label{eq:3.19}
\end{equation}
while the same quantity in the source region is represented
in the complete form
\begin{eqnarray}    \fl
\it\Phi_1({\bi r}_{\rm t},z) &=
\it\Phi_a({\bi r}_{\rm t},z) + \it\Phi_b({\bi r}_{\rm t},z) =
\sum_l A_l(z)\,\it\Phi_l({\bi r}_{\rm t},z) +
\it\Phi_b({\bi r}_{\rm t},z)      \nonumber\\
\fl  &=\, \sum_l A_l(z)\,
\hat{\!\it\Phi}_l({\bi r}_{\rm t})\,{\rm e}^{-\,\gamma_l z}\! +
\it\Phi_b({\bi r}_{\rm t},z) \equiv
\sum_l a_l(z)\,\hat{\!\it\Phi}_l({\bi r}_{\rm t}) +
\it\Phi_b({\bi r}_{\rm t},z)
\label{eq:3.20}
\end{eqnarray}
involving the modal expansion $\it\Phi_a$ and the orthogonal complement
$\it\Phi_b$. The latter is true for all physical quantities except the
curl fields and quasi-static potentials for which, according to
equations~(\ref{eq:1.4}) and (\ref{eq:1.5}), ${\bi E}_{{\rm c}b}=
{\bi H}_{{\rm c}b}^{\rm p}= 0$ and $\varphi_b= \psi_b= 0$.

The separation of potential fields causes the surface boundary conditions
(formerly written for the total fields in the form of relations~(4.6) and
(4.7) in~[1]) to be reformulatted. Inside the source region,
besides the external bulk currents ${\bi J}_{\rm b}^{\rm e,m}$ and charges
$\rho_{\rm b}^{\rm e,m}$, there are the surface sources, which are
located along the contour $L_{\rm s}$ and given in the form of:

(a) the {\it current sheet\/} with the electric and magnetic surface current
densities ${\bi J}_{\rm s}^{\rm e}$ and ${\bi J}_{\rm s}^{\rm m}$ which
results in discontinuity in the tangential components of the curl magnetic
and electric fields ${\bi n}_{\rm s}\times{\bi H}_{{\rm c}1}^{\rm p}$ and
${\bi n}_{\rm s}\times{\bi E}_{{\rm c}1}$, respectively;

(b) the {\it charge sheet\/} with the electric and magnetic surface charge
densities $\rho_{\rm s}^{\rm e}$ and $\rho_{\rm s}^{\rm m}$ which results
in discontinuity in the normal component of the electric and magnetic
inductions ${\bi n}_{\rm s}\bdot{\bi D}_1^{\rm p}$ and
${\bi n}_{\rm s}\bdot{\bi B}_1$, respectively;

(c) the {\it dipole (double charge) sheet\/} with the electric and magnetic
surface dipole densities $\bfeta_{\rm s}^{\rm e}$ and
$\bfeta_{\rm s}^{\rm m}$ which results in discontinuity in the
quasi-static electric and magnetic potentials ${\varphi}_1$ and ${\psi}_1$,
respectively.

The corresponding boundary conditions written in pairs for the electric
and magnetic sources have the following form~[14,\,20]:

(a) for the current sheets
\begin{eqnarray}
{\bi n}_{\rm s}^+ \times\,{\bi E}_{{\rm c}1}^+            & + &\,
{\bi n}_{\rm s}^-\!\times{\bi E}_{{\rm c}1}^- =
-\,{\bi J}_{\rm s}^{\rm m}
\label{eq:3.21}\\             [.2cm]
{\bi n}_{\rm s}^+\!\times{\bi H}_{{\rm c}1}^{{\rm p}+}    & + &\,
{\bi n}_{\rm s}^-\times\,{\bi H}_{{\rm c}1}^{{\rm p}-} =\;
{\bi J}_{\rm s}^{\rm e}
\label{eq:3.22}
\end{eqnarray}

(b) for the charge sheets
\begin{eqnarray}
{\bi n}_{\rm s}^+\!\bdot{\bi D}_1^{{\rm p}+}              & + &\,
{\bi n}_{\rm s}^-\!\bdot{\bi D}_1^{{\rm p}-} \!=\,\rho_{\rm s}^{\rm e}
\label{eq:3.23}\\             [.2cm]
{\bi n}_{\rm s}^+\!\bdot\,{\bi B}_1^+                     & + &\,
{\bi n}_{\rm s}^-\!\bdot{\bi B}_1^- \,=\,\rho_{\rm s}^{\rm m}
\label{eq:3.24}
\end{eqnarray}

(c) for the dipole sheets
\begin{eqnarray}
{\bi n}_{\rm s}^+\,\varphi_1^+             & + &\,
{\bi n}_{\rm s}^-\,\varphi_1^- \,=\,
{1\over\epsilon_0}\;\bfeta_{\rm s}^{\rm e}
\label{eq:3.25}\\
{\bi n}_{\rm s}^+\,\psi_1^+                & + &\,
{\bi n}_{\rm s}^-\,\psi_1^- \,=\,
{1\over\mu_0}\;\bfeta_{\rm s}^{\rm m} .
\label{eq:3.26}
\end{eqnarray}

Here, all the quantities with superscripts $^\pm$ are values taken
at points of the contour $L_s$ lying on its different sides marked
by the inward unit normal vector ${\bi n}_s^\pm$.

For deriving the equations of mode excitation it is necessary to
integrate the reciprocity relation~(\ref{eq:3.6}) (with replacing
subscripts 2 by $k$) over the cross section $S$ of a waveguide by using
relation~(2.25) in~[1], with the result depending on whether the
potential fields are separated.

Without separating the potential fields (when ${\bi E}_1= {\bi E}_a +
{\bi E}_b$ and ${\bi H}_1^{\rm p}= {\bi H}_a^{\rm p}+ {\bi H}_b^{\rm p}$
in accordance with equation~(\ref{eq:3.20})), the substitution of
equation~(\ref{eq:3.8}) (with changing subscripts $2\!\to\!k$) into the
integral relation~(2.25) of~[1] yields
\[       \fl
\int_S \bnabla\bdot{\bi S}_{1k}^{\rm EM}\,\d S \,=\,
{\partial\over\partial z} \int_S
\Bigl( {\bi E}_a\times{\bi H}_k^{{\rm p}*} +
{\bi E}_k^*\times{\bi H}_a^{\rm p} \Bigr) \bdot{\bi z}_0\,\d S
\]
\[     \fl
+ \int_{L_{\rm b}} \!\!{\bi n}_{\rm b}\bdot
\Bigl( {\bi E}_b\times{\bi H}_k^{{\rm p}*} +
{\bi E}_k^*\times{\bi H}_b^{\rm p} \Bigr) \d l \,-
\int_{L_{\rm s}} \biggl[ \,
{\bi n}_{\rm s}^+\!\bdot \Bigl( {\bi E}_1\times{\bi H}_k^{{\rm p}*} +
{\bi E}_k^*\times{\bi H}_1^{\rm p} \Bigr)^{\!+}
\]
\[      \fl
+\;{\bi n}_{\rm s}^-\!\bdot \Bigl( {\bi E}_1\times{\bi H}_k^{{\rm p}*} +
{\bi E}_k^*\times{\bi H}_1^{\rm p} \Bigr)^{\!-} \biggr] \d l \;\equiv\;
{\partial\over\partial z} \int_S
{\bi S}_{1k}^{\rm EM}\!\bdot{\bi z}_0\,\d S
\]
\begin{equation}       \fl
+ \int_{L_{\rm s}} \Bigl( {\bi J}_{\rm s}^{\rm e}\bdot{\bi E}_k^* +
{\bi J}_{\rm s}^{\rm m}\bdot{\bi H}_k^{{\rm p}*} \Bigr) \d l +
\int_{L_{\rm b}} \Bigl( {\bi J}_{\rm s,ef}^{\rm e}\bdot{\bi E}_k^* +
{\bi J}_{\rm s,ef}^{\rm m}\bdot{\bi H}_k^{{\rm p}*} \Bigr) \d l
\label{eq:3.27}
\end{equation}
where
\begin{equation}
{\bi S}_{1k}^{\rm EM} = {\bi E}_1\times{\bi H}_k^{{\rm p}*} \,+\,
{\bi E}_k^*\times{\bi H}_1^{\rm p} \,.
\label{eq:3.28}          \\[.2cm]
\end{equation}
Here we have used: (i) the boundary conditions~(4.6) and (4.7)
of paper~[1] with the real surface currents ${\bi J}_{\rm s}^{\rm e}$ and
${\bi J}_{\rm s}^{\rm m}$ given on the contour $L_{\rm s}$ with the unit
normal vectors ${\bi n}_{\rm s}^\pm$,\, (ii)~the effective surface currents
${\bi J}_{\rm s,ef}^{\rm e}= -\,{\bi n}_{\rm b}\times{\bi H}_b^{\rm p}$
and ${\bi J}_{\rm s,ef}^{\rm m}= {\bi n}_{\rm b}\times{\bi E}_b$ defined
on the boundary contour $L_{\rm b}$ of the bulk source area $S_{\rm b}$
with the outward unit normal vector ${\bi n}_{\rm b}$ (see~(4.24)
-- (4.28) in~[1]).

With separating the potential fields (when ${\bi E}_{{\rm c}b}=
{\bi H}_{{\rm c}b}^{\rm p}= 0$ and $\varphi_b=\psi_b= 0$ in accordance
with equations~(\ref{eq:1.4}) and (\ref{eq:1.5})), the substitution of
(\ref{eq:3.13}) (with changing subscripts $2\!\to\!k$) into the
integral relation~(2.25) of~[1] yields
\[         \fl
\int_S \bnabla\bdot{\bi S}_{1k}^{\rm EM}\,\d S\,=\,
{\partial\over\partial z} \int_S \, \biggl[
\Bigl( {\bi E}_{{\rm c}1}\times
{\bi H}_{{\rm c}k}^{{\rm p}*} +
{\bi E}_{{\rm c}k}^*\times
{\bi H}_{{\rm c}1}^{\rm p} \Bigr)    \\[-0.15cm]
\]
\[          \fl
+\,\Bigl( \varphi_1(\i\omega{\bi D}_k^{\rm p})^* \!+
\varphi_k^*(\i\omega{\bi D}_1^{\rm p}) \Bigr) +
\Bigl( \psi_1(\i\omega{\bi B}_k)^* \!+
\psi_k^*(\i\omega{\bi B}_1) \Bigr) \biggr] \!\bdot{\bi z}_0\,\d S
\]
\[          \fl
- \int_{L_{\rm s}} \biggl[ \,
{\bi n}_{\rm s}^+\!\bdot \Bigl( {\bi E}_{{\rm c}1}\times
{\bi H}_{{\rm c}k}^{{\rm p}*} \!+
{\bi E}_{{\rm c}k}^*\times{\bi H}_{{\rm c}1}^{\rm p} \Bigr)^{\!+} \!+
{\bi n}_{\rm s}^-\!\bdot \Bigl( {\bi E}_{{\rm c}1}\times
{\bi H}_{{\rm c}k}^{{\rm p}*} \!+
{\bi E}_{{\rm c}k}^*\times
{\bi H}_{{\rm c}1}^{\rm p} \Bigr)^{\!-} \biggr] \d l
\]
\[          \fl
-\! \int_{L_{\rm s}} \biggl[ \,
{\bi n}_{\rm s}^+\!\bdot \Bigl( \varphi_1(\i\omega{\bi D}_k^{\rm p})^* \!+
\varphi_k^*(\i\omega{\bi D}_1^{\rm p}) \Bigr)^{\!+} \!+
{\bi n}_{\rm s}^-\!\bdot \Bigl( \varphi_1(\i\omega{\bi D}_k^{\rm p})^* \!+
\varphi_k^*(\i\omega{\bi D}_1^{\rm p}) \Bigr)^{\!-} \biggr] \d l
\]
\[          \fl
-\! \int_{L_{\rm s}} \biggl[ \,
{\bi n}_{\rm s}^+\!\bdot \Bigl( \psi_1(\i\omega{\bi B}_k)^* \!+
\psi_k^*(\i\omega{\bi B}_1) \Bigr)^{\!+} \!+
{\bi n}_{\rm s}^-\!\bdot \Bigl( \psi_1(\i\omega{\bi B}_k)^* \!+
\psi_k^*(\i\omega{\bi B}_1) \Bigr)^{\!-} \biggr] \d l \,.
\]

The use of the boundary conditions~(\ref{eq:3.21}) -- (\ref{eq:3.26})
for rearranging line integrals in the last formula gives
\[        \fl
\int_S \bnabla\bdot{\bi S}_{1k}^{\rm EM}\,\d S\,=\,
{\partial\over\partial z} \int_S {\bi S}_{1k}^{\rm EM}\!\bdot
{\bi z}_0\,\d S \,+
\int_{L_{\rm s}} \Bigl( {\bi J}_{\rm s}^{\rm e}\bdot
{\bi E}_{{\rm c}k}^* +
{\bi J}_{\rm s}^{\rm m}\bdot{\bi H}_{{\rm c}k}^{{\rm p}*} \Bigr) \d l
\]
\begin{equation}           \fl
- \int_{L_{\rm s}} \biggl[
\Bigl( \i\omega\rho_{\rm s}^{\rm e}\,\varphi_k^* +
\i\omega\rho_{\rm s}^{\rm m}\,\psi_k^* \Bigr) +
\biggl( \, {\bfeta_{\rm s}^{\rm e}\over\epsilon_0}\bdot
(\i\omega{\bi D}_k^{\rm p})^* \!+
{\bfeta_{\rm s}^{\rm m}\over\mu_0}\bdot
(\i\omega{\bi B}_k)^* \biggr) \biggr] \d l
\label{eq:3.29}
\end{equation}
where
\begin{eqnarray}
{\bi S}_{1k}^{\rm EM} &=
\Bigl( {\bi E}_{{\rm c}1}\times{\bi H}_{{\rm c}k}^{{\rm p}*} +
{\bi E}_{{\rm c}k}^*\times{\bi H}_{{\rm c}1}^{\rm p} \Bigr) +
\Bigl( \varphi_1(\i\omega{\bi D}_k^{\rm p})^* +
\varphi_k^*(\i\omega{\bi D}_1^{\rm p}) \Bigr)   \nonumber\\
&+ \,\Bigl( \psi_1(\i\omega{\bi B}_k)^* +
\psi_k^*(\i\omega{\bi B}_1) \Bigr) \,.
\label{eq:3.30}
\end{eqnarray}

It should be noted that, as distinct from equation~(\ref{eq:3.27}),
expression (\ref{eq:3.29}) does not contain the effective surface sources
like ${\bi J}_{\rm s,ef}^{\rm e,m}$ since the curl fields~(\ref{eq:1.4})
and the quasi-static potentials~(\ref{eq:1.5}) have no orthogonal
complements.

Substitution of equation~(\ref{eq:3.9}) (with changing subscripts
$2\!\to\!k$) into the integral relation~(2.25) of~[1] yields
\begin{equation}
\int_S \bnabla\bdot{\bi S}_{1k}^{\rm PM}\,\d S\,=\,
{\partial\over\partial z} \int_S
{\bi S}_{1k}^{\rm PM}\!\bdot{\bi z}_0\,\d S
\label{eq:3.31}
\end{equation}
where
\begin{eqnarray}
{\bi S}_{1k}^{\rm PM} =
&- \Bigl( \bar{\bi T}_1^{\scriptscriptstyle\Sigma}\bdot{\bi U}_k^* +
\bar{\bi T}_k^{\Sigma*}\bdot{\bi U}_1 \Bigr) +
\Bigl( \bar{\bi V}_1^{\rm m}\bdot{\bi J}_k^{{\rm m}*} +
\bar{\bi V}_k^{{\rm m}*}\bdot{\bi J}_1^{\rm m} \Bigr)   \nonumber\\
&+\, \Bigl( \bar{\bi V}_1^{\rm e}\bdot{\bi J}_k^{{\rm e}*} +
\bar{\bi V}_k^{{\rm e}*}\!\bdot{\bi J}_1^{\rm e} \Bigr)
\label{eq:3.32}
\end{eqnarray}

Integrating the reciprocity relation~(\ref{eq:3.6}) (with changing
subscripts $2\!\to\!k$) over the cross section $S$ of a waveguiding
structure and using~(\ref{eq:3.27}), (\ref{eq:3.28}) or
(\ref{eq:3.29}), (\ref{eq:3.30}) along with (\ref{eq:3.31}) and
(\ref{eq:3.32}) give the following result:
\begin{equation}
{\d P_{1k}(z)\over\d z}\,+\,Q_{1k}(z)\,=\,R_{1k}(z)
\label{eq:3.34}
\end{equation}
where
\begin{equation}
P_{1k} = P_{1k}^{\rm EM} + P_{1k}^{\rm PM} =
\int_S \Bigl( {\bi S}_{1k}^{\rm EM} +\,
{\bi S}_{1k}^{\rm PM} \Bigr) \!\bdot{\bi z}_0\,\d S
\label{eq:3.35}
\end{equation}
\begin{equation}
Q_{1k} = \int_S q_{1k}\,\d S
\label{eq:3.36}
\end{equation}
\begin{equation}
R_{1k} =\, R_{1k}^{({\rm b})} \,+\, R_{1k}^{({\rm s})} =
\int_{S_{\rm b}} \!r_{1k}^{({\rm b})}\,\d S \,+
\int_{L_{\rm s}} \!r_{1k}^{({\rm s})}\,\d l \,.
\label{eq:3.37}
\end{equation}

The loss term $Q_{1k}$ has the universal form determined by formula
(\ref{eq:3.10}) (with changing subscripts $2\!\to\!k$) for $q_{1k}$.
In contrast, the expressions for $P_{1k}$ and $R_{1k}$ are obtained
different depending on whether the potential fields are separated.

Without separating the potential fields, the power term~(\ref{eq:3.35})
involves ${\bi S}_{1k}^{\rm EM}$ and ${\bi S}_{1k}^{\rm PM}$ given by
(\ref{eq:3.28}) and (\ref{eq:3.32}), respectively. The bulk
excitation term determined by~(\ref{eq:3.11}) (with
\,${\bi J}_{{\rm b}2}^{\rm e,m}= 0$ and changing subscripts
$2\!\to\!k$) is equal to
\begin{equation}
R_{1k}^{({\rm b})}\equiv \int_{S_{\rm b}} \!r_{1k}^{({\rm b})}\,\d S =
- \int_{S_{\rm b}} \Bigl( {\bi J}_{\rm b}^{\rm e}\bdot{\bi E}_k^* \,+\,
{\bi J}_{\rm b}^{\rm m}\bdot{\bi H}_k^{{\rm p}*} \Bigr) \d S
\label{eq:3.38}
\end{equation}
and the surface excitation term follows from equation~(\ref{eq:3.27})
in the form
\begin{eqnarray}
R_{1k}^{({\rm s})}\equiv
\int_{L_{\rm s}+L_{\rm b}} \!r_{1k}^{({\rm s})}\,\d l =
&- \int_{L_{\rm s}} \Bigl(
{\bi J}_{\rm s}^{\rm e}\bdot{\bi E}_k^* \,+\,
{\bi J}_{\rm s}^{\rm m}\bdot{\bi H}_k^{{\rm p}*} \Bigr) \d l  \nonumber\\
&- \int_{L_{\rm b}} \Bigl(
{\bi J}_{\rm s,ef}^{\rm e}\bdot{\bi E}_k^* \,+\,
{\bi J}_{\rm s,ef}^{\rm m}\bdot{\bi H}_k^{{\rm p}*} \Bigr) \d l .
\label{eq:3.39}
\end{eqnarray}

With separating the potential fields, the quantity ${\bi S}_{1k}^{\rm EM}$
and ${\bi S}_{1k}^{\rm PM}$ determining the power term~(\ref{eq:3.35})
are given by (\ref{eq:3.30}) and (\ref{eq:3.32}), respectively.
The bulk excitation term determined by equations~(\ref{eq:3.14}) (with
\,${\bi J}_{{\rm b}2}^{\rm e,m}= \rho_{{\rm b}2}^{\rm e,m}= 0$ and
changing subscripts $2\!\to\!k$) is equal to
\begin{eqnarray}
R_{1k}^{({\rm b})}\equiv
\int_{S_{\rm b}} \!r_{1k}^{({\rm b})}\,\d S =
\int_{S_{\rm b}} \biggl[
&- \Bigl( {\bi J}_{\rm b}^{\rm e}\bdot{\bi E}_{{\rm c}k}^* \,+\,
{\bi J}_{\rm b}^{\rm m}\bdot
{\bi H}_{{\rm c}k}^{{\rm p}*} \Bigr) \nonumber\\
&+ \Bigl( \i\omega\rho_{\rm b}^{\rm e}\,\varphi_k^*  \,+\,
\i\omega\rho_{\rm b}^{\rm m}\,\psi_k^* \Bigr) \biggr] \d S
\label{eq:3.40}
\end{eqnarray}
and the surface excitation term, according to equation~(\ref{eq:3.29}),
take the form
\[             \fl
R_{1k}^{({\rm s})}\equiv \int_{L_{\rm s}} \!r_{1k}^{({\rm s})}\,\d l =
\int_{L_{\rm s}} \biggl[ -\,\Bigl(
{\bi J}_{\rm s}^{\rm e}\bdot{\bi E}_{{\rm c}k}^* \,+\,
{\bi J}_{\rm s}^{\rm m}\bdot{\bi H}_{{\rm c}k}^{{\rm p}*} \Bigr)
\]
\begin{equation}    \fl \qquad
+\,\Bigl( \i\omega\rho_{\rm s}^{\rm e}\,\varphi_k^* +
\i\omega\rho_{\rm s}^{\rm m}\,\psi_k^* \Bigr) \,+\,
\biggl( \, {\bfeta_{\rm s}^{\rm e}\over\epsilon_0}\bdot
(\i\omega{\bi D}_k^{\rm p})^* +\,
{\bfeta_{\rm s}^{\rm m}\over\mu_0}\bdot
(\i\omega{\bi B}_k)^* \biggr) \biggr] \d l \,.
\label{eq:3.41}
\end{equation}

The quadratic (power) quantities $P_{1k}$ and $Q_{1k}$ in the form of
(\ref{eq:3.35}) and (\ref{eq:3.36}) are constructed of the
linear quantities: (i)~$\it{\Phi}_k$ of the form~(\ref{eq:3.19}) and
(ii)~$\it{\Phi}_1=\it{\Phi}_a+\it{\Phi}_b$ of the form~(\ref{eq:3.20})
involving the modal expansion $\it{\Phi}_a=\sum_l A_l\it{\Phi}_l$ and
the orthogonal complement $\it{\Phi}_b$ (the latter is absent for the
curl fields and quasi-static potentials). On this basis, we can rewrite
(\ref{eq:3.35}) and (\ref{eq:3.36}) as
\begin{eqnarray}     \fl
P_{1k} = P_{ak} + P_{bk}     \;\qquad
P_{ak} = P_{ak}^{\rm EM} + P_{ak}^{\rm PM} =
\int_S \Bigl( {\bi S}_{ak}^{\rm EM} +\,
{\bi S}_{ak}^{\rm PM} \Bigr) \!\bdot{\bi z}_0\,\d S
\label{eq:3.42}\\          [.15cm]
\fl
Q_{1k} = Q_{ak} + Q_{bk}        \qquad
Q_{ak} = \int_S q_{ak}\,\d S \,.
\label{eq:3.43}
\end{eqnarray}
Here the subscripts $a$ and $b$ correspond to using the modal expansions
$\it{\Phi}_a$ and the orthogonal complements $\it{\Phi}_b$ to construct the
appropriate quadratic quantities;\, therewith $P_{bk}= P_{bk}^{\rm EM}+
P_{bk}^{\rm PM}\equiv P_{bk}^{\rm PM}$ since always $P_{bk}^{\rm EM}\!= 0$
by virtue of the fact that without separating potential fields
${\bi S}_{bk}^{\rm EM}\!\bdot{\bi z}_0\!= 0$ because of ${\bi E}_b =
{\bi z}_0 E_b$ and ${\bi H}_b^{\rm p}= {\bi z}_0 H_b^{\rm p}$,\, whereas
with separating them ${\bi S}_{bk}^{\rm EM}\equiv 0$ because of
(\ref{eq:1.4}) and (\ref{eq:1.5}).

The basic property of the orthogonal complement to be (quasi-)orthogonal
in power sense with respect to the fields of every $k$th eigenmode,
assures the following quasi-orthogonality relation for lossy systems:
\begin{equation}
{\d P_{bk}(z)\over\d z}\,+\,Q_{bk}(z) =\,0
\label{eq:3.44}
\end{equation}
which is written in agreement with the mode quasi-orthogonality relation
(\ref{eq:3.15}).\, Hence, with allowing for (\ref{eq:3.42}) --
(\ref{eq:3.44}) formula~(\ref{eq:3.34}) takes the form
\begin{equation}
{\d P_{ak}(z)\over\d z}\,+\,Q_{ak}(z) =\,R_{1k}(z)
\label{eq:3.45}
\end{equation}
involving solely the modal expansions with no orthogonal complements.

Substitution of the modal expansions into expressions~(\ref{eq:3.42}) and
(\ref{eq:3.43}) for $P_{ak}$ and $Q_{ak}$ allows us to reveal their
longitudinal dependence (without an explicit writing of transverse
coordinates):
\begin{eqnarray}     \fl
P_{ak}(z) &= P_{ak}^{\rm EM}(z) + P_{ak}^{\rm PM}(z) \equiv
\int_S \Bigl( {\bi S}_{ak}^{\rm EM}(z) \,+\,
{\bi S}_{ak}^{\rm PM}(z) \Bigr) \!\bdot{\bi z}_0\,\d S    \nonumber\\
\fl &= \sum_l A_l(z)\! \int_S \Bigl( {\bi S}_{lk}^{\rm EM} +\,
{\bi S}_{lk}^{\rm PM} \Bigr) \!\bdot{\bi z}_0\,\d S =
\sum_l A_l(z)\! \int_S \Bigl( {\bi S}_{kl}^{{\rm EM}*} \!+\,
{\bi S}_{kl}^{{\rm PM}*} \Bigr) \!\bdot{\bi z}_0\,\d S    \nonumber\\
\fl &= \sum_l \Bigl( N_{kl}^{\rm EM} \!+ N_{kl}^{\rm PM} \Bigr)
A_l(z)\,{\rm e}^{-\,(\gamma_k^* +\gamma_l) z} \equiv
\Bigl( \,\sum_l  N_{kl}\,A_l(z)\,{\rm e}^{-\,\gamma_l z} \Bigr)\,
{\rm e}^{-\,\gamma_k^* z}
\label{eq:3.46}
\end{eqnarray}
\begin{eqnarray}      \fl
Q_{ak}(z) &= \int_S q_{ak}(z)\,\d S=
\sum_l A_l(z)\,\int_S q_{lk}\,\d S =        \nonumber\\
\fl &= \sum_l A_l(z)\, \int_S q_{kl}^*\,\d S \equiv
\Bigl( \,\sum_l  M_{kl}\,A_l(z)\,{\rm e}^{-\,\gamma_l z} \Bigr) \,
{\rm e}^{-\,\gamma_k^* z} \,.
\label{eq:3.47}
\end{eqnarray}

By inserting equations~(\ref{eq:3.46}) and (\ref{eq:3.47}) in relation
(\ref{eq:3.45}) and representing the exciting integrals $R_{1k}$ in
the form
\[
R_{1k} = R_{1k}^{({\rm b})} + R_{1k}^{({\rm s})} =
\Bigl( R_k^{({\rm b})}+ R_k^{({\rm s})} \Bigr) \,
{\rm e}^{-\,\gamma_k^* z}\equiv R_k\,{\rm e}^{-\,\gamma_k^* z}
\]
involving the wave factor $\exp(-\,\gamma_k^* z)$ explicitly, we obtain
\begin{equation}         \fl
\sum_l \biggl\{ N_{kl}\,{\d A_l\over\d z} -
\Bigl[\, (\gamma_k^* +\gamma_l) N_{kl}- M_{kl}\, \Bigr] A_l
\biggr\} \,{\rm e}^{-\,\gamma_l z} =\,
R_k \equiv\,R_k^{({\rm b})} +\, R_k^{({\rm s})} \,.
\label{eq:3.48}
\end{equation}

The square bracket in equation~(\ref{eq:3.48}) vanishes owing to the
quasi-orthogonality relation~(\ref{eq:3.18}) and the required
equations of mode excitation take the following~form:\\
(i) for the excitation amplitudes $A_l(z)$
\begin{equation}
\sum_l N_{kl}\,{\d A_l(z)\over\d z}\,{\rm e}^{-\,\gamma_l z} =\,
R_k^{({\rm b})}(z) +
R_k^{({\rm s})}(z) ,   \qquad  k= 1,2,\ldots
\label{eq:3.49}
\end{equation}
(ii) for the mode amplitudes \,$a_l(z)= A_l(z)\,{\rm e}^{-\gamma_l z}$
\begin{equation}
\sum_l N_{kl} \biggl(
{\d a_l(z)\over\d z} + \gamma_l a_l(z) \biggr) =\,
R_k^{({\rm b})}(z) + R_k^{({\rm s})}(z) , \quad  k= 1,2,\ldots
\label{eq:3.50}
\end{equation}

These equations have the general structure applicable for both cases of
separating and not separating the potential fields. The only distinction
between them consists in different forms of the normalizing coefficients
$N_{kl}= N_{kl}^{\rm EM}+ N_{kl}^{\rm PM}$ (electromagnetic and
polarized-medium) and the exciting integrals $R_k= R_k^{({\rm b})}+
R_k^{({\rm s})}$ (bulk and surface) which depend on whether the potential
fields are separated:\\
(a) {\it without separating the potential fields\/} (when equations
(\ref{eq:3.38}) and (\ref{eq:3.39}) are~valid)
\[        \fl
N_{kl} = N_{kl}^{\rm EM} + N_{kl}^{\rm PM} = \int_S \Bigl(
\hat{\!\bi E}_k^*\times\hat{\!\bi H}_l^{\rm p} +
\hat{\!\bi E}_l\times\hat{\!\bi H}_k^{{\rm p}*}
\Bigr) \!\bdot {\bi z}_0\,\d S +
\int_S \, \biggl[
-\Bigl(\!\!\!\!\!\hat{\,\,\;\;\bar{\bi T}_k^{\Sigma*}}\bdot
\hat{\bi U}_l +
\!\!\!\hat{\,\,\;\bar{\bi T}_l^{\scriptscriptstyle\Sigma}}\bdot
\hat{\bi U}_k^* \Bigr)
\]
\begin{equation}      
+\; \Bigl(\!\!\!\!\!\!\hat{\,\;\;\;\bar{\bi V}_k^{{\rm m}*}}\!\bdot
\hat{\!\bi J}_l^{\rm m}+
\!\!\!\!\hat{\,\;\;\bar{\bi V}_l^{\rm m}}\!\bdot
\hat{\!\bi J}_k^{{\rm m}*} \Bigr) \,+\,
\Bigl( \!\!\!\!\hat{\,\;\;\bar{\bi V}_k^{{\rm e}*}}\!\bdot
\hat{\!\bi J}_l^{\rm e} +
\!\!\hat{\,\;\bar{\bi V}_l^{\rm e}}\!\bdot
\hat{\!\bi J}_k^{{\rm e}*} \Bigr)
\biggr] \!\bdot{\bi z}_0\,\d S
\label{eq:3.51}
\end{equation}
\[      \fl
R_k = R_k^{({\rm b})} + R_k^{({\rm s})} =
- \int_{S_{\rm b}} \Bigl( {\bi J}_{\rm b}^{\rm e}\bdot
\hat{\!\bi E}_k^* +
{\bi J}_{\rm b}^{\rm m}\bdot
\hat{\!\bi H}_k^{{\rm p}*} \Bigr) \d S \,-
\int_{L_{\rm s}} \Bigl(
{\bi J}_{\rm s}^{\rm e}\bdot\hat{\!\bi E}_k^* +
{\bi J}_{\rm s}^{\rm m}\bdot
\hat{\!\bi H}_k^{{\rm p}*} \Bigr) \d l
\]
\begin{equation}
-\int_{L_{\rm b}} \Bigl(
{\bi J}_{\rm s,ef}^{\rm e}\bdot\hat{\!\bi E}_k^* +
{\bi J}_{\rm s,ef}^{\rm m}\bdot\hat{\!\bi H}_k^{{\rm p}*} \Bigr) \d l
\label{eq:3.52}
\end{equation}
involving necessarily the effective surface currents
${\bi J}_{\rm s,ef}^{\rm e}$ and ${\bi J}_{\rm s,ef}^{\rm m}$ which
were developed in~[1] (see formulae (4.24 -- 4.32)).\\
(b) {\it with separating the potential fields\/} (when equations
(\ref{eq:3.40}) and (\ref{eq:3.41}) are~valid)
\[         \fl
N_{kl} =\,N_{kl}^{\rm EM} + \,N_{kl}^{\rm PM} = \int_S \, \biggl[
\Bigl( \hat{\!\bi E}_{{\rm c}k}^*\times
\hat{\!\bi H}_{{\rm c}l}^{\rm p} +\,
\hat{\!\bi E}_{{\rm c}l}\times
\hat{\!\bi H}_{{\rm c}k}^{{\rm p}*} \Bigr) +
\Bigl(  \hat{\varphi}_k^*\,(\i\omega\,\hat{\!\bi D}_l^{\rm p}) +
\hat{\varphi}_l\,(\i\omega\,\hat{\!\bi D}_k^{\rm p})^* \Bigr)
\]
\[
+ \Bigl(  \hat{\psi}_k^*\,(\i\omega\,\hat{\!\bi B}_l) +
\hat{\psi}_l\,(\i\omega\,\hat{\!\bi B}_k)^* \Bigr)
\biggr] \!\bdot{\bi z}_0 \d S
+ \int_S \, \biggl[
-\Bigl(\!\!\!\!\!\hat{\,\,\;\;\bar{\bi T}_k^{\Sigma*}}\bdot
\hat{\bi U}_l +
\!\!\!\hat{\,\,\;\bar{\bi T}_l^{\scriptscriptstyle\Sigma}}\bdot
\hat{\bi U}_k^* \Bigr)
\]
\begin{equation}
+ \Bigl(\!\!\!\!\!\!\hat{\,\;\;\;\bar{\bi V}_k^{{\rm m}*}}\!\bdot
\hat{\!\bi J}_l^{\rm m}+
\!\!\!\!\hat{\,\;\;\bar{\bi V}_l^{\rm m}}\!\bdot
\hat{\!\bi J}_k^{{\rm m}*} \Bigr) \,+\,
\Bigl( \!\!\!\!\hat{\,\;\;\bar{\bi V}_k^{{\rm e}*}}\!\bdot
\hat{\!\bi J}_l^{\rm e} +
\!\!\hat{\,\;\bar{\bi V}_l^{\rm e}}\!\bdot
\hat{\!\bi J}_k^{{\rm e}*} \Bigr)
\biggr] \!\bdot{\bi z}_0 \d S
\label{eq:3.53}
\end{equation}
\[         \fl
R_k = R_k^{({\rm b})} + R_k^{({\rm s})} \!=
\int_{S_{\rm b}} \biggl[ - \Bigl( {\bi J}_{\rm b}^{\rm e}\bdot
\hat{\!\bi E}_{{\rm c}k}^* +
{\bi J}_{\rm b}^{\rm m}\bdot\hat{\!\bi H}_{{\rm c}k}^{{\rm p}*} \Bigr) +
\Bigl( \i\omega\rho_{\rm b}^{\rm e}\,\hat{\varphi}_k^* +
\i\omega\rho_{\rm b}^{\rm m}\,\hat{\psi}_k^*\Bigr)\biggr] \d S
\]
\[
+ \int_{L_{\rm s}} \biggl[ - \Bigl(
{\bi J}_{\rm s}^{\rm e}\bdot\hat{\!\bi E}_{{\rm c}k}^* +
{\bi J}_{\rm s}^{\rm m}\bdot\hat{\!\bi H}_{{\rm c}k}^{{\rm p}*} \Bigr) +
\Bigl( \i\omega\rho_{\rm s}^{\rm e}\,\hat{\varphi}_k^* +
\i\omega\rho_{\rm s}^{\rm m}\,\hat{\psi}_k^* \Bigr)
\]
\begin{equation}
+\,\biggl( {\bfeta_{\rm s}^{\rm e}\over\epsilon_0}\bdot
(\i\omega\,\hat{\!\bi D}_k^{\rm p})^* +
{\bfeta_{\rm s}^{\rm m}\over\mu_0}\bdot
(\i\omega\,\hat{\!\bi B}_k)^* \biggr) \biggr] \d l
\label{eq:3.54}     \\[.1cm]
\end{equation}
involving no effective surface currents, unlike the previous case.

All the quantities $N_{kl}$ and $R_k$ involve the cross-section
eigenfunctions (marked by the hat sign above them) and the $z$-dependence
of the exciting integrals $R_k(z)$ is due to that of the external bulk
and surface sources.

The separation of potential fields has revealed a fine structure
of the interaction between the external sources and the mode eigenfields
(curl and potential), which is demonstrated by relation~(\ref{eq:3.54}):\,
the currents (electric ${\bi J}_{\rm b,s}^{\rm e}$ and magnetic
${\bi J}_{\rm b,s}^{\rm m}$) interact with the curl fields (electric
$\hat{\!\bi E}_{{\rm c}k}$ and magnetic $\hat{\!\bi H}_{{\rm c}k}^{\rm p}$),
whereas the charges (electric $\rho_{\rm b,s}^{\rm e}$ and magnetic
$\rho_{\rm b,s}^{\rm m}$) interact with the quasi-static potentials
(electric $\varphi_k$ and magnetic $\psi_k$). Also, there is the
interaction of the displacement currents (electric
$\i\omega\,\hat{\!\bi D}_k^{\rm p}$ and magnetic
$\i\omega\,\hat{\!\bi B}_k$) with the double charge (dipole)
layers (electric $\bfeta_{\rm s}^{\rm e}$ and magnetic
$\bfeta_{\rm s}^{\rm m}$), if any. Usually, the latter do not exist
in real physical situations, but can be introduced by the equivalence
principle as equivalent surface sources.

It should be mentioned that the generalized theory of guided-wave
interaction, allowing for the potental fields, much like the theory
elaborated here, was first developed by the author~[22]. Both results
are in good agreement only for lossless waveguiding structures,
although the problem of the orthogonal complements went unnoticed then.
Allowance for losses was made on the basis of the bi-orthogonality
relation (instead of the quasi-orthogonality relation, as it is done
here) which was obtained by introducing a subsidiary boundary-value
problem (called the associated problem) to change artificially
the sign of a loss parameter. In so doing, the eigenmode norms for
lossy waveguides have no power meaning and the results of mode
excitation pose serious difficulties in physical interpretation.

\section{General conclusions on the excitation theory for BAM and SDAM
                      waveguides}
\label{sec:4}

The final equations (\ref{eq:3.49}) or (\ref{eq:3.50}) of the waveguide
excitation theory developed are completely identical, by their structure,
with equations~(5.47) and (5.48) obtained in~[1] for the waveguides
with bianisotropic media. They constitute an infinite set of coupled
differential equations of the first order in the desired modal amplitudes
$A_k$ or $a_k$ excited by a given distribution of the external sources
(bulk and surface) which enter into the exciting integrals
$R_k^{({\rm b})}$ and $R_k^{({\rm s})}$ defined by
formulae~(\ref{eq:3.52}) and (\ref{eq:3.54}). This set of excitation
equations can be rewritten in the matrix-operator form
\begin{equation}
\bar{\bi N}\bdot{\bi Z}(z) =\,{\bi R}(z)
\label{eq:4.1}
\end{equation}
where the matrix-operator $\bar{\bi N}$ formed from the normalizing
coefficients is hermitian $(N_{kl}= N_{lk}^*)$ and the column-vector
${\bi R}$ has $R_k$ as components. The column-vector ${\bi Z}$ is
composed from the elements $Z_k = \d a_k/\d z +\gamma_k a_k =
(\d A_k/\d z)\exp(-\gamma_k z)$.

Generally, the coupled equations developed hold true for dissipative
systems since the coupling coefficient $N_{kl}$ defined by
equation~(\ref{eq:3.51}) or (\ref{eq:3.53}) determines, according to
(\ref{eq:3.16}), the cross-power flow $P_{kl}$ for any one of mode
pairs, which is the case for lossy waveguides. It~is evident that these
equations remain true for nondissipative systems as a special~case.

For lossless waveguiding structures, the general quasi-orthogonality
relation~(\ref{eq:3.18}) turns into equation~(3.10) of paper~[1]
to produce there the orthonormalization relations (3.18) and (3.24)
for the {\it active\/} and {\it reactive\/} modes, respectively.
They can be written jointly in the combined form
\begin{equation}
N_{kl} =\, \left\{ \begin{array}{cl}
   N_k\delta_{kl} & \qquad\mbox{-- for \,\,active \,\,modes}  \\[.15cm]
N_k\delta_{{\tilde k}l} & \qquad\mbox{-- for reactive modes}
\end{array}  \right.
\label{eq:4.3}
\end{equation}
where the norms $N_k$ possess such a property that for an active mode
it is real-valued, whereas for a reactive mode it is complex-valued
and $N_k\!= N_{\tilde k}^*$ (see section 3.2 in~[1]). Substitution
of equation~(\ref{eq:4.3}) into (\ref{eq:4.1}) yields
\begin{equation}
Z_k =\, \left\{ \begin{array}{cl}
                  R_k / N_k & \qquad\mbox
                              {-- for \,\,active \,\,modes}  \\[.15cm]
R_{\tilde k} / N_{\tilde k} & \qquad\mbox
                              {-- for reactive modes}
\end{array}  \right.
\label{eq:4.4}
\end{equation}

As follows from the orthogonality relations~(\ref{eq:4.3}), every
eigenmode of a lossless waveguide is orthogonal to all the other
modes, except for the only one in combination with fields of which
it forms its own norm: the active $k$-mode is non-orthogonal to itself
but the reactive $k$-mode is non-orthogonal to its own twin-conjugate
$\tilde k$-mode for which $\gamma_k+\gamma_{\tilde k}^*= 0$. The
fields of such a mode, non-orthogonal to the $k$-mode, enter into the
exciting integrals and determine the power of interaction with external
sources, so as to supply the given $k$-mode independently of the others
(see section 3.2 in~[1]). This is true only if the exciting
sources are fixed, which lies outside the context of self-consistent
treatment usually applied in practice.

The independent excitation of every mode by given sources described
by (\ref{eq:4.4}) is inherent only in lossless waveguides.
For a lossy waveguide this is not the case.
Even for the fixed exciting sources there is a dissipative coupling
among modes expressed by non-zero off-diagonal elements $N_{kl}$ of the
normalizing coefficient matrix $\bar{\bi N}$. Such a dissipative coupling
implies that, unlike lossless waveguides, the given external sources
excite the total set of eigenmodes as a whole rather than every mode as
a single. It is very important to realize that this mode coupling of
dissipative character is apparent since it exists only inside sources.
Outside them even the presence of nonzero off-diagonal coefficients
$N_{kl}$ leaves all the eigenmodes uncoupled. Indeed, for the
source-free region the right-hand side of equation~(\ref{eq:4.1}) vanishes
$(R_k= 0)$ and, \,if $N_{kl}\ne 0$,\, from here it follows that
$Z_k= 0$\, or \,$A_k(z)$ = constant.

Hence, the eigenmodes of a lossy waveguide, being linearly independent
solutions to the appropriate boundary-value problem, remain uncoupled
outside the source region. Under these conditions, any eigenmode has
the constant value of amplitude that was gained from sources at the exit
boundary of their existence region and propagates along the waveguide
without coupling to other modes. However, the picture of power transfer
is more involved. Every $k$-mode, besides the {\it self-power\/} flow
$P_k\equiv P_{kk}$, also carries the {\it cross-power\/} flows $P_{kl}$ in
pairs, together with the other $l$-modes, which were also excited inside
the source region and outside have constant amplitudes. According to the
quasi-orthogonality relation in the power form~(\ref{eq:3.15}), any one
of mode pairs has a certain value of the cross-power flow $P_{kl}$, as well
as the cross-power loss $Q_{kl}$. This fact is a physical manifestation of
the power non-orthogonality (called the {\it quasi-orthogonality\/}) among
eigenmodes outside the source region. It is the existence of $P_{kl}$
proportional to the normalizing coefficients $N_{kl}$ that results in
the apparent coupling of dissipative character among modes inside the
source region, which is described by equations~(\ref{eq:3.49}) or
(\ref{eq:3.50}) with the exciting sources on the right considered
as~fixed.

At first glance, it may seem that the infinite set of coupled equations
like~(\ref{eq:4.1}) obtained for lossy systems reduces to certain
mathematical difficulties which are absent for lossless systems described
by the uncoupled equations like~(\ref{eq:4.4}). However, the difference
between them disappears in the self-consistent formulation of a wave
problem. In this case, the external sources appearing in $R_k$ themselves
can be represented as series expansions in terms of eigenmodes of another
waveguide, which makes these sources exciting for the waveguide
under consideration. Then, the uncoupled equations for single modes, for
example~(\ref{eq:4.4}), turn into an infinite set of coupled equations
(see Refs.\,[13,\,22,\,23] where this technique is described). Solving
such coupled equations for lossless systems differs little in computational
complexity from an analogous solution for lossy systems. In both cases,
to obtain the total solution of coupled equations, allowing for interaction
between all modes, is unrealizable in practice. Usually, the major
effect of interaction is determined by coupling among finite number of
modes. Strongly interacting modes can be separated by the coupled-mode
technique~[24], which enables an approximate solution to be obtained with
a precision sufficient for practical applications.

Thus, we have developed a unified treatment of the electrodynamic theory
of guided-wave excitation by external sources, applied to any waveguide
with composite and multilayered structures involving complex media, with
both bi-anisotropic and space-dispersive properties.

\appendix
\section{Derivation of the generalized reciprocity relation for
                        space-dispersive active media}
\label{app:A}

\subsection{Contribution from piezoelectrically-elastic properties
                                   of a medium}
\label{app:A1}

Let us rewrite equations~(\ref{eq:2.11}) -- (\ref{eq:2.16}) for the first
system with subscript~1\,:
\begin{eqnarray}
T_{1,ij} & = & c_{ijkl}\,S_{1,kl} - e_{kij}\,E_{1,k}
\label{eq:A1}\\
D_{1,k}  & = & \,e_{kij}\,S_{1,ij}\;+\;\epsilon_{ik}\,E_{1,i}
\label{eq:A2}
\end{eqnarray}
\begin{equation}
\i\omega\rho_{\rm m} U_{1,i} =\,
{\partial T_{1,ij}^{\scriptscriptstyle\Sigma}\over\partial r_j}\,-\,
\tau_{ij}^{-1}\rho_{\rm m} U_{1,j}
\label{eq:A3}
\end{equation}
\begin{equation}
T_{1,ij}^{\scriptscriptstyle\Sigma} =\,
T_{1,ij} + T_{1,ij}^{\rm fr}\,, \quad
T_{1,ij}^{\rm fr} =\,\i\omega\eta_{ijkl}\,S_{1,kl}
\label{eq:A4}
\end{equation}
\begin{equation}
\i\omega S_{1,ij} =\,
{1\over2}\, \biggl( {\partial U_{1,i}\over\partial r_j}\,+\,
{\partial U_{1,j}\over\partial r_i} \biggr) .
\label{eq:A5}
\end{equation}

For the second system, the similar equations are obtained from equations
(\ref{eq:A1}) through (\ref{eq:A5}) by taking complex conjugation and
replacing subscripts 1 with~2 (both have the small-signal meaning). The
equations obtained in such a way are numbered as (A$.1^\prime$)
through (A$.5^\prime$), but will not be written explicitly for the sake
of brevity.

In order to calculate the term
\[
\i\omega \Bigl( {\bi P}_1\bdot{\bi E}_2^* -
{\bi P}_2^*\bdot{\bi E}_1 \Bigr) \equiv\,
\i\omega \Bigl( {\bi D}_1\bdot{\bi E}_2^* -
{\bi D}_2^*\bdot{\bi E}_1 \Bigr)
\]
in the right-hand side of equation~(\ref{eq:3.5}) it is necessary to:
multiply (A.1) and (A$.1^\prime$) by $-\i\omega S_{2,ij}^*$ and
\,$\i\omega S_{1,ij}$, respectively; multiply (A.2) and (A$.2^\prime$) by
$-\i\omega E_{2,k}^*$ and $\i\omega E_{1,k}$, respectively;\, and to add
all the results.\, Then
\begin{equation}
\i\omega \Bigl( {\bi P}_1\bdot{\bi E}_2^* -
{\bi P}_2^*\bdot{\bi E}_1 \Bigr) =\,
\i\omega \Bigl( T_{1,ij}\,S_{2,ij}^* - T_{2,ij}^*\,S_{1,ij} \Bigr)
\label{eq:A6}
\end{equation}
where the relations \,$\epsilon_{ik}=\epsilon_{ki}$\, and
\,$c_{ijkl}=c_{klij}$\, have been applied.\\
Multiplying (A.5) by $T_{2,ij}^*$\, and (A$.5^\prime$)
by $T_{1,ij}$  gives the following
\[        \fl
\i\omega \Bigl( T_{1,ij}\,S_{2,ij}^* - T_{2,ij}^*\,S_{1,ij} \Bigr) =
-\, \biggl( T_{1,ij}\,{\partial U_{2,i}^*\over\partial r_j} +
T_{2,ij}^*\,{\partial U_{1,i}\over\partial r_j} \biggr)
\]
\[        \fl
= -\,\bnabla\bdot \Bigl(
\bar{\bi T}_1\bdot{\bi U}_2^* \,+\,
\bar{\bi T}_2^*\bdot{\bi U}_1 \Bigr) \,+\,
\biggl( U_{1,i}\,{\partial T_{2,ij}^*\over\partial r_j}\,+\,
U_{2,i}^*\,{\partial T_{1,ij}\over\partial r_j}  \biggr)
\]
\[            \fl
= -\,\bnabla\bdot \Bigl(
\bar{\bi T}_1^{\scriptscriptstyle\Sigma}\bdot{\bi U}_2^* +
\bar{\bi T}_2^{\Sigma*}\bdot{\bi U}_1 \Bigr) +
\biggl( U_{1,i}\,{\partial T_{2,ij}^{\Sigma*}\over\partial r_j} +
U_{2,i}^*\,{\partial T_{1,ij}^{\scriptscriptstyle\Sigma}\over\partial r_j}
\biggr)
\]
\begin{equation}
+\,\biggl( T_{1,ij}^{\rm fr}\,{\partial U_{2,i}^*\over\partial r_j} +
T_{2,ij}^{\rm fr*}\,{\partial U_{1,i}\over\partial r_j} \biggr) .
\label{eq:A7}
\end{equation}
Multiplying (A.3) by $U_{2,i}^*$\, and (A$.3^\prime$)
by $U_{1,i}$ gives the following
\[
\biggl( U_{1,i}\,{\partial T_{2,ij}^{\Sigma*}\over\partial r_j} \,+\,
U_{2,i}^*\,
{\partial T_{1,ij}^{\scriptscriptstyle\Sigma}\over\partial r_j}\biggr)=
\rho_{\rm m}\tau_{ij}^{-1} \Bigl(
U_{1,i}\,U_{2,j}^* +\, U_{2,i}^*\,U_{1,j} \Bigr)
\]
\begin{equation}
=\,2\,\rho_{\rm m}\tau_{ij}^{-1} U_{1,i}\,U_{2,j}^* \equiv
2\,\rho_{\rm m}{\bi U}_2^* \bdot\,
\bar{\!\btau}^{-1}\!\bdot{\bi U}_1
\label{eq:A8}
\end{equation}
where the last equality is obtained by using the relation $\tau_{ij}^{-1}=
\tau_{ji}^{-1}$.

To calculate the last term in the right-hand side of~(\ref{eq:A7})
\,it is necessary to multiply (A.4) and (A$.4^\prime$) by
${\partial U_{2,i}^*/\partial r_j}$ and ${\partial U_{1,i}/\partial r_j}$,
\,respectively,\, and to add the results.\, Then the use of (A.5)
and (A$.5^\prime$) yields
\[       \fl
\biggl( T_{1,ij}^{\rm fr}\,{\partial U_{2,i}^*\over\partial r_j} \,+\,
T_{2,ij}^{fr*}\,{\partial U_{1,i}\over\partial r_j}  \biggr) =\,
\i\omega\eta_{ijkl} \biggl(
S_{1,kl}\,{\partial U_{2,i}^*\over\partial r_j} -
S_{2,kl}^*\,{\partial U_{1,i}\over\partial r_j} \biggr)
\]
\[            \fl
= \i\omega\,{\eta_{ijkl}\over2} \Biggl[
S_{1,kl} \biggl( {\partial U_{2,i}^*\over\partial r_j} +
{\partial U_{2,j}^*\over\partial r_i}  \biggr) -
S_{2,kl}^* \biggl( {\partial U_{1,i}\over\partial r_j} +
{\partial U_{1,j}\over\partial r_i}  \biggr) \Biggr]
\]
\begin{equation}      \fl
=\, \omega^2\eta_{ijkl} \Bigl(
S_{1,ij}\,S_{2,kl}^* + S_{2,ij}^*\,S_{1,kl} \Bigr) =
2\,\omega^2\eta_{ijkl}\,S_{1,ij}\,S_{2,kl}^* \equiv
2\,\omega^2\,\bar{\bi S}_2^* \!:\,
\bar{\bar{\!\bfeta}}: \bar{\bi S}_1
\label{eq:A9}
\end{equation}
where the relations \,$\eta_{ijkl}= \eta_{klij}$\, and
\,$\eta_{ijkl}= \eta_{jikl}$ have been applied. \\[.2cm]
By collecting formulae~(\ref{eq:A6}) -- (\ref{eq:A9}) we finally
obtain
\[
\i\omega \Bigl( {\bi P}_1\bdot{\bi E}_2^* -
{\bi P}_2^*\bdot{\bi E}_1 \Bigr) =
-\,\bnabla\bdot \Bigl(
\bar{\bi T}_1^{\scriptscriptstyle\Sigma}\bdot{\bi U}_2^* \,+\,
\bar{\bi T}_2^{\Sigma*}\bdot{\bi U}_1 \Bigr)
\]
\begin{equation}
+\;2\,\omega^2\,\bar{\bi S}_2^* \!:\,
\bar{\bar{\!\bfeta}}: \bar{\bi S}_1 \,+\,
2\,\rho_{\rm m}{\bi U}_2^* \bdot\,
\bar{\!\btau}^{-1}\!\bdot{\bi U}_1 \,.
\label{eq:A10}
\end{equation}

\subsection{Contribution from ferrimagnetic properties of a medium}
\label{app:A2}

For ferrimagnetic magnetic media it is necessary to calculate the term
\[
\i\omega\mu_0 \Bigl( {\bi M}_1\bdot{\bi H}_2^* -
{\bi M}_2^*\bdot{\bi H}_1 \Bigr) \equiv\,
\i\omega \Bigl( {\bi B}_1\bdot{\bi H}_2^* -
{\bi B}_2^*\bdot{\bi H}_1 \Bigr)
\]
on the right-hand side of~(\ref{eq:3.5}). By using
(\ref{eq:2.21}) -- (\ref{eq:2.33}) we write the linearized
equation of motion for the first system marked by subscript~1:
\begin{eqnarray}
\i\omega{\bi M}_1 = &- \gamma\mu_0 \Bigl[\,
{\bi M}_1\times{\bi H}_0\,+\,{\bi M}_0\times{\bi H}_1 -
{\bi M}_0\times(\;\bar{\bcal N}\bdot{\bi M}_1)   \nonumber\\
&+ \lambda_{\rm ex}\,{\bi M}_0\times
\bnabla^2{\bi M}_1 \Bigr] \,+\,
i\nu_{\scriptscriptstyle M}
{\omega\over\omega_{\scriptscriptstyle M}} \biggl(\,
{{\bi M}_0\over M_0}\times{\bi M}_1 \!\biggr)
\label{eq:A11}
\end{eqnarray}
where \,$\nu_{\scriptscriptstyle M}= \alpha\omega_{\scriptscriptstyle M}=
\alpha\gamma\mu_0 M_0$\, is the magnetic relaxation frequency.

The similar equation for the second system marked by subscript 2 in place
of 1 is obtained from equation~(\ref{eq:A11}) by taking complex conjugation
and numbered as (A$.11^\prime$) (without its explicit writing for brevity).

Vector-multiplying equation~(\ref{eq:A11}) by ${\bi M}_2^*$ and
taking account of the equalities ${\bi M}_{1,2}\bdot{\bi H}_0=
{\bi M}_{1,2}\bdot{\bi M}_0= 0$ valid for small signals we obtain
\begin{eqnarray}   \fl
\i\omega \Bigl( {\bi M}_1\times{\bi M}_2^* \Bigr) =
& \gamma\mu_0 \biggl[
{\bi M}_0 \Bigl( {\bi M}_2^*\bdot{\bi H}_1 \Bigr) -
{\bi H}_0 \Bigl( {\bi M}_2^*\bdot{\bi M}_1 \Bigr) -
{\bi M}_0 \Bigl( {\bi M}_2^*\bdot\,\bar{\bcal N}
\bdot{\bi M}_1 \Bigr)         \nonumber\\
&+ \lambda_{\rm ex}\,{\bi M}_0 \Bigl(
{\bi M}_2^*\bdot\bnabla^2{\bi M}_1 \Bigr) \biggr] \,-\,
i\nu_{\scriptscriptstyle M}{\omega\over\omega_{\scriptscriptstyle M}}
{{\bi M}_0\over M_0} \Bigl( {\bi M_2^*}\bdot{\bi M}_1 \Bigr) \,.
\label{eq:A12}
\end{eqnarray}
The similar equation obtained in a such way from equation\,
(A$.11^\prime$) is
\begin{eqnarray}      \fl
\i\omega \Bigl( {\bi M}_2^*\times{\bi M}_1 \Bigr) =
&- \gamma\mu_0 \biggl[
{\bi M}_0 \Bigl( {\bi M}_1\!\bdot{\bi H}_2^* \Bigr) -
{\bi H}_0 \Bigl( {\bi M}_1\!\bdot{\bi M}_2^* \Bigr) -
{\bi M}_0 \Bigl( {\bi M}_1\!\bdot\,\bar{\bcal N}
\bdot{\bi M}_2^* \Bigr)        \nonumber\\
&+ \lambda_{\rm ex}\,{\bi M}_0 \Bigl(
{\bi M}_1\!\bdot\bnabla^2{\bi M}_2^* \Bigr) \biggr] \,-\,
i\nu_{\scriptscriptstyle M}{\omega\over\omega_{\scriptscriptstyle M}}
{{\bi M}_0\over M_0} \Bigl( {\bi M_1}\!\bdot{\bi M}_2^* \Bigr) \,.
\label{eq:A13}
\end{eqnarray}

By adding these equations with using the vector-dyadic identity\,
$\bnabla\bdot(\bnabla{\bi A}\bdot{\bi B})={\bi B}\bdot\bnabla^2{\bi A}+
(\bnabla\times{\bi A})\bdot(\bnabla\times{\bi B})+
\bnabla{\bi A}:\bnabla{\bi B}$\, and allowing for the symmetry of
tensor $\bar{\bcal N}$ we finally obtain
\[
\i\omega\mu_0 \Bigl( {\bi M}_1\bdot{\bi H}_2^* -
{\bi M}_2^*\bdot{\bi H}_1 \Bigr)\,=
\]
\begin{equation}
=\,\bnabla\bdot \Bigl(
\bar{\bi V}_1^{\rm m}\bdot{\bi J}_2^{{\rm m}*}+
\bar{\bi V}_2^{{\rm m}*}\bdot{\bi J}_1^{\rm m} \Bigr)
+\,2\,\nu_{\scriptscriptstyle M}\mu_0 \biggl(
{\omega\over\omega_{\scriptscriptstyle M}} \biggr)^{\!2}
\Bigl( {\bi M_1}\bdot{\bi M}_2^* \Bigr)
\label{eq:A14}
\end{equation}
where the magnetization current \,${\bi J}_{1,2}=
\i\omega\mu_0{\bi M}_{1,2}$\, and the effective magnetic (exchange)
potential \,$\bar{\bi V}_{1,2}^{\rm m} =
-\lambda_{\rm ex}\bnabla{\bi M}_{1,2}$\, of a
ferrimagnetic medium have been used.

\subsection{Contribution of drifting charge carriers in a medium}
\label{app:A3}

For plasmas with drifting charge carriers, it is necessary to calculate
the term
\[       \fl
\i\omega \Bigl( {\bi p}_1\bdot{\bi E}_2^{\prime*} -
{\bi p}_2^*\bdot{\bi E}_1^\prime \Bigr) \equiv\,
\i\omega \Bigl( {\bi p}_1 \bdot
({\bi E}_2^* + {\bi v}_0\times{\bi B}_2^*) -
{\bi p}_2^*\bdot({\bi E}_1 + {\bi v}_0\times{\bi B}_1) \Bigr)
\]
on the right-hand side of~(\ref{eq:3.5}). To this end, we apply
(\ref{eq:2.42}) and (\ref{eq:2.43}) for the first system with
subscript~1 written in the form
\[      \fl
\i\omega{\bi v}_1 +\,
({\bi v}_0\bdot\bnabla){\bi v}_1\,=
\]
\[         \fl
=\,{e\over m} \Bigl[ \,
{\bi E}_1\,+\,({\bi r}_1\bdot\bnabla){\bi E}_0\,+\,
{\bi v}_1\times{\bi B}_0\,+\,{\bi v}_0\times{\bi B}_1\,+\,
{\bi v}_0\times({\bi r}_1\bdot
\bnabla){\bi B}_0\,\Bigr]
\]
\begin{equation}     \fl
+\;{v_{\scriptscriptstyle T}^2\over\rho_0}\, \Bigl[ \,
\rho_0\bnabla(\bnabla\bdot{\bi r}_1)\,+\,
\bnabla{\bi r}_1\bdot\bnabla\rho_0\,
\Bigr]\,-\, {{\bi v}_1\over\tau_0}\,+\,{{\bi v}_0\over\tau_0}\,{\tau_1 +
({\bi r}_1\bdot\bnabla)\tau_0\over\tau_0}
\label{eq:A15}
\end{equation}
\begin{equation}
{\bi v}_1 =\,\i\omega{\bi r}_1 +\,
({\bi v}_0\bdot\bnabla){\bi r}_1
\label{eq:A16}
\end{equation}
and the similar equations for the second system with subscript~2
in place of~1 obtained from~(\ref{eq:A15}) and (\ref{eq:A16})
by taking complex conjugation and numbered as (A$.15^\prime$) and
(A$.16^\prime$) (without explicit writing for~brevity).

Multiplying~(\ref{eq:A15}) and (A$.15^\prime$) by
$-\,\i\omega(m/e)\rho_0{\bi r}_2^*$ and $\i\omega(m/e)\rho_0{\bi r}_1$,
respectively,\, and adding the results with a combination of terms give
\[    \fl
\i\omega \Bigl( {\bi p}_1\bdot{\bi E}_2^{\prime*} -
{\bi p}_2^*\bdot{\bi E}_1^\prime \Bigr) \,=\,
-\,\i\omega\rho_0\,{m\over e}\, \biggl\{
\i\omega \Bigl( {\bi r}_2^*\bdot{\bi v}_1 +
{\bi r}_1\bdot{\bi v}_2^* \Bigr)
\]
\[        \fl
+ \Bigl[ {\bi r}_2^*\bdot({\bi v}_0\bdot\bnabla){\bi v}_1 -
{\bi r}_1\bdot({\bi v}_0\bdot\bnabla){\bi v}_2^*
\Bigr]  \biggr\} \,+\,
\i\omega\rho_0 \Bigl[ \,
{\bi r}_2^*\bdot({\bi r}_1\bdot\bnabla){\bi E}_0 -
{\bi r}_1\bdot({\bi r}_2^*\bdot\bnabla){\bi E}_0\,\Bigr]
\]
\[        \fl
+\,\i\omega\rho_0  \biggl\{  \Bigl[
{\bi r}_2^*\bdot ({\bi v}_1\times{\bi B}_0) -
{\bi r}_1\bdot ({\bi v}_2^*\times{\bi B}_0) \Bigr] +
\Bigl[ {\bi r}_2^*\bdot \Bigl( {\bi v}_0\times({\bi r}_1\bdot
\bnabla){\bi B}_0 \Bigr)
\]
\[          \fl
-\, {\bi r}_1\bdot \Bigl( {\bi v}_0\times({\bi r}_2^*\bdot
\bnabla){\bi B}_0 \Bigr) \Bigr] \biggr\} \,+\,
\i\omega\,v_{\scriptscriptstyle T}^2 {m\over e} \biggl\{  \Bigl[
\rho_0{\bi r}_2^*\bdot\bnabla(\bnabla\bdot{\bi r}_1) -
\rho_0{\bi r}_1\bdot\bnabla(\bnabla\bdot{\bi r}_2^*) \Bigr] +
\]
\[         \fl
+\,\Bigl[ {\bi r}_2^*\bdot(\bnabla{\bi r}_1\bdot\bnabla\rho_0) -
{\bi r}_1\bdot(\bnabla{\bi r}_2^*\bdot\bnabla\rho_0) \,\Bigr]  \biggr\}
\]
\begin{equation}      \fl
+\,\i\omega\rho_0\,{1\over\tau_0}{m\over e}\,  \biggl[ \,
{\bi r}_1 \bdot \biggl(  {\bi v_2^*} - {\bi v}_0\, {\tau_2^* +
{\bi r}_2^*\!\bdot\!\bnabla\tau_0\over\tau_0} \biggr) -
{\bi r}_2^* \bdot \biggl(  {\bi v_1} - {\bi v}_0\, {\tau_1 +
{\bi r}_1\!\bdot\!\bnabla\tau_0\over\tau_0}
\biggr) \biggr]\,.
\label{eq:A17}
\end{equation}

Let us transform the right-hand side of~(\ref{eq:A17}), with
the last terms already in the desired form.

The first term is transformed by using
$\bnabla\bdot{\bi J}_0= 0$, and the result of
multiplying~(\ref{eq:A16}) and (A$.16^\prime$) by
$-\,\i\omega(m/e)\rho_0{\bi v}_2^*$\, and \,$\i\omega(m/e)\rho_0{\bi v}_1$,
respectively,\, into the following form
\[         \fl
-\,\i\omega\rho_0\,{m\over e}\, \biggl\{
\i\omega \Bigl( {\bi r}_2^*\bdot{\bi v}_1 +
{\bi r}_1\bdot{\bi v}_2^* \Bigr) +\, \Bigl[
{\bi r}_2^*\bdot({\bi v}_0\bdot\bnabla){\bi v}_1 -
{\bi r}_1\bdot({\bi v}_0\bdot\bnabla){\bi v}_2^*
\Bigr]  \biggr\}
\]
\begin{equation}   \fl
=\,\bnabla\bdot \biggl[ \,
\i\omega\,{m\over e}\,{\bi v}_0 \Bigl(
{\bi p}_1\bdot{\bi v}_2^* - {\bi p}_2^*\bdot{\bi v}_1 \Bigr) \biggr] \,.
\label{eq:A18}
\end{equation}

The second term vanishes after using the following vector-dyadic identity
\begin{equation}
({\bi A}\times{\bi B})\bdot(\bnabla\times{\bi C})=
{\bi B}\bdot({\bi A}\bdot\bnabla{\bi C}) -
{\bi A}\bdot({\bi B}\bdot\bnabla{\bi C})
\label{eq:id}
\end{equation}
with ${\bi A}\!=\!{\bi r}_1$,\,\, ${\bi B}={\bi r}_2^*$ and
${\bi C}={\bi E}_0$, since $\bnabla\times{\bi E}_0= 0$.

The third term is rearranged by employing~(\ref{eq:A16}),\,
(A$.16^\prime$),\, $\bnabla\bdot{\bi B}_0= 0$,\,\,
$\bnabla\bdot{\bi J}_0= 0$,\, and the identity
\[
{\bi A}\times({\bi B}\bdot\bnabla{\bi C})-
{\bi B}\times({\bi A}\bdot\bnabla{\bi C})=
\bnabla{\bi C}\bdot({\bi B}\times{\bi A})-
(\bnabla\bdot{\bi C})({\bi B}\times{\bi A})
\]
with ${\bi A}={\bi r}_1,\; {\bi B}={\bi r}_2^*$ and ${\bi C}={\bi B}_0$.
Therefore,
\[         \fl
\i\omega\rho_0  \biggl\{  \Bigl[
{\bi r}_2^*\bdot ({\bi v}_1\times{\bi B}_0) -
{\bi r}_1\bdot ({\bi v}_2^*\times{\bi B}_0) \Bigr] +
\Bigl[ {\bi r}_2^*\bdot \Bigl( {\bi v}_0\times({\bi r}_1\bdot
\bnabla){\bi B}_0 \Bigr)
\]
\[      \fl
-\,{\bi r}_1\bdot \Bigl( {\bi v}_0\times({\bi r}_2^*\bdot
\bnabla){\bi B}_0 \Bigr) \Bigr] \biggr\} =
\i\omega\rho_0 \Bigl[ \, {\bi B}_0\bdot \Bigl(
{\bi r}_2^*\times{\bi v}_1 - {\bi r}_1\times{\bi v}_2^* \Bigr) \,+\,
{\bi v}_0\bdot \Bigl(
\bnabla{\bi B}_0\bdot({\bi r}_2^*\times{\bi r}_1) \Bigr)  \Bigr]
\]
\begin{equation}     \fl
=-\,\bnabla\bdot \Bigl[ \,
\i\omega{\bi v}_0 \Bigl(
{\bi p}_1\bdot({\bi r}_2^*\times{\bi B}_0)/2 -
{\bi p}_2^*\bdot({\bi r}_1\times{\bi B}_0)/2 \Bigr) \Bigr] \,.
\label{eq:A19}
\end{equation}

The fourth term uses identity~(\ref{eq:id}) with ${\bi A}={\bi r}_1,\;
{\bi B}={\bi r}_2^*,\; {\bi C}=\bnabla\rho_0$ and the fact that
$\bnabla\times\bnabla\rho_0= 0$,\, to be transformed
into the following~form
\[         \fl
\i\omega v_{\scriptscriptstyle T}^2{m\over e} \biggl\{  \Bigl[
\rho_0{\bi r}_2^*\bdot\bnabla
(\bnabla\bdot{\bi r}_1) -
\rho_0{\bi r}_1\bdot\bnabla
(\bnabla\bdot{\bi r}_2^*) \Bigr] +
\Bigl[ {\bi r}_2^*\bdot
(\bnabla{\bi r}_1\bdot\bnabla\rho_0)
\]
\[             \fl
- {\bi r}_1\bdot(\bnabla{\bi r}_2^*\bdot\bnabla\rho_0) \Bigr] \biggr\}=
\i\omega v_{\scriptscriptstyle T}^2\,{m\over e}\, \biggl[ \,
\bnabla\bdot \Bigl(
\rho_0{\bi r}_2^*\,\bnabla\bdot{\bi r}_1 -
\rho_0{\bi r}_1\,\bnabla\bdot  {\bi r}_2^* \Bigr)
\]
\[         \fl
+\,\Bigl( \bnabla\bdot{\bi p}_1\,
\bnabla\bdot{\bi r}_2^* -
\bnabla\bdot{\bi p}_2^*\,
\bnabla\bdot{\bi r}_1  \Bigr) \,+\,
\Bigl( {\bi r}_2^*\bdot\bnabla{\bi r}_1 -
{\bi r}_1\bdot\bnabla{\bi r}_2^* \Bigr) \bdot
\bnabla\rho_0\, \biggr]
\]
\begin{equation}        \fl
=\,\bnabla\bdot \biggl[ \,
\i\omega\,{m\over e}\,{v_{\scriptscriptstyle T}^2\over\rho_0}\, \Bigl(
{\bi p}_2^*\,\bnabla\bdot{\bi p}_1 -
{\bi p}_1\,\bnabla\bdot{\bi p}_2^* \Bigr) \biggr] \equiv
\bnabla\bdot \Bigl(
V_1^{\rm th}\,{\bi J}_2^{{\rm e}*}\,+\,V_2^{et*}\,{\bi J}_1^{\rm e} \Bigr)
\label{eq:A20}
\end{equation}
where the thermal (diffusion) potential \,$V_{1,2}^{\rm th}=
(k_BT/e)(\rho_{1,2}/\rho_0)$ has been used.

Substituting (\ref{eq:A18}) -- (\ref{eq:A20}) into
(\ref{eq:A17}), we finally obtain
\[        \fl
\i\omega \Bigl( {\bi p}_1\bdot{\bi E}_2^{\prime*} -\,
{\bi p}_2^*\bdot{\bi E}_1^\prime \Bigr) \,=\,
\bnabla\bdot \Bigl(
\bar{\bi V}_1^{\rm e}\bdot{\bi J}_2^{{\rm e}*}\,+\,
\bar{\bi V}_2^{{\rm e}*}\!\bdot{\bi J}_1^{\rm e} \Bigr)
\]
\begin{equation}      \fl
+\,{1\over\mu_{\rm e}}\,\Biggl[
\biggl( {\bi v}_1 - {\bi v}_0\, {\tau_1 +{\bi r}_1\bdot
\bnabla\tau_0\over\tau_0} \biggr)\bdot
{\bi J}_2^{{\rm e}*} +
\biggl( {\bi v}_2^* - {\bi v}_0\, {\tau_2^* +{\bi r}_2^*\bdot
\bnabla\tau_0\over\tau_0}\biggr)\bdot{\bi J}_1^{\rm e}
\, \Biggr]
\label{eq:A21}
\end{equation}
where the electronic polarization current \,${\bi J}_{1,2}^{\rm e}=
\i\omega{\bi p}_{1,2}$\, and the effective electronic potential\,
$\bar{\bi V}_{1,2}^{\rm e}= \bar{\bi V}_{1,2}^{\rm ek} +
V_{1,2}^{\rm th}\,\bar{\bi I}= (m/e)\,({\bi v}_0{\bi v}_{1,2}^{\rm p} +
(v_{\scriptscriptstyle T}^2/\rho_0)\,\rho_{1,2}\,\bar{\bi I}\,)$\,
have been introduced.

\end{document}